\documentclass[a4paper, 11pt]{article}
\usepackage[utf8]{inputenc}
\usepackage[T1]{fontenc}
\usepackage[english]{babel}
\usepackage[width=160mm,height=247mm]{geometry}
\usepackage{amsmath, amsfonts, amssymb, bm}
\usepackage[pdftex]{graphicx}
\usepackage{lmodern, indentfirst, cite, float}
\usepackage[squaren]{SIunits}
\usepackage[position=top,labelfont={color=blue,bf,sf}]{subfig}
\usepackage[textfont={footnotesize},labelfont={color=blue,bf,sf},labelsep=endash]{caption}
\usepackage[colorlinks=true,linkcolor=blue,citecolor=blue,urlcolor=blue]{hyperref}

\usepackage[affil-sl]{authblk}
\title{\textbf{\Large Electron flow in Monolayer Molybdenum Disulfide Quantum Dot with Magnetic Flux}}
\author[1]{Abdelhadi Belouad\thanks{\href{mailto:belabdelhadi@gmail.com}{belabdelhadi@gmail.com}}}
\author[1,2]{Abdellatif Kamal\thanks{\href{mailto:abdellatif.kamal@univh2c.ma}{
abdellatif.kamal@univh2c.ma}}}
\author[1,3]{Rachid Houça}
\author[1]{El Bouâzzaoui Choubabi}
\affil[1]{L.P.M.C. Laboratory, Theoretical Physics Group, Faculty of Sciences, Choua\"ib Doukkali University, PO Box 20, 24000 El Jadida, Morocco}
\affil[1,2]{ISPS2I Laboratory, National Higher School of Arts and Crafts (ENSAM), Hassan II University, 20670, Casablanca, Morocco}
\affil[3]{Equipe de Physique Théorique et Hautes Energies, Faculté des Sciences, université Ibn Zohr, PO Box 8106, Agadir, Maroc}
\date{\today}
\begin{document}
%========================================================
\begin{titlepage}
	\newgeometry{width=175mm, height=247mm}
    \maketitle
    \thispagestyle{empty}
    \vspace{3cm}
    \begin{abstract}
		We study the Dirac electron scattering problem on a potential barrier in a circular quantum dot of Monolayer Molybdenum Disulfide quantum dot $\mathrm{MoS_2}$ subjected to magnetic flux. By solving Dirac's equation, we formulate analytical expressions for the eigenstates, scattering coefficients, scattering efficiency, and the reflected current's radial component. We show that the scattering coefficients, the scattering efficiency, and the radial component of the reflected current depend explicitly on the magnetic flux. The magnetic flux may cause a slight shift in the oscillation position of the scattering coefficients. For scattering efficiency, magnetic flux may also lead to a slight shift in their oscillation position and an increase in amplitude when the magnetic flux decreases.
	\end{abstract}
	\vspace{02cm}
\end{titlepage}
\restoregeometry
%========================================================
\section{Introduction}
%========================================================
The discovery of graphene, the first single-atom thick material, in 2004 \cite{ Novoselov04, Novoselov041}, has accelerated research into atomically thin two-dimensional (2D) materials. The capacity to control isolated single atomic layers and reassemble them layer-by-layer in a precise sequence to build heterostructures opens up endless possibilities for applications \cite{Geim13,Wang12,Yazyev15}. Semiconducting dichalcogenides are attractive compounds in this regard since they can be readily exfoliated and have a suitably narrow gap both in bulk and as a single layer.
Monolayer molybdenum disulfide (MoS2) has recently attracted attention in this type of 2D dichalcogenides systems for combining electron mobility comparable to graphene devices with a limited energy gap \cite{Radisavljevic01}. Monolayer MoS2 possesses a direct gap,  which is an indirect gap semiconductor \cite{Geim13,Mak10}, making it particularly intriguing for optoelectronics. Another intriguing aspect is that the electrical properties appear to be very sensitive to external pressure \cite{Nayak12}, strain \cite{Kang12,Conley13}, and temperature \cite{Radisavljevic13}, which influence the gap and, under certain conditions, can also produce an insulator/metal transition. Furthermore, the lack of lattice inversion symmetry combined with spin-orbit coupling (SOC) leads to coupled spin and valley physics in monolayers of MoS2 and other group VI dichalcogenides \cite{Xiao12}, allowing control of spin and valley in these materials \cite{Kadantsev12}. Because of their unique band structure, monolayers of MoS2 have been proposed for a range of nanoelectronics applications \cite{Wang12}, including valleytronics, spintronics, optoelectronics, and room temperature transistor devices \cite{Radisavljevic01}.

Due to their intriguing features, including as optical properties, electrochemical and catalytic activities, two-dimensional (2D) transition-metal dichalcogenide nanomaterials have piqued the interest of many researchers in recent years. As a result, various potential applications in sensing, photocatalysis, energy, healthcare, and other related domains have been proven \cite{Ping17,Singh17,Tan15,Gong17,Yadav19}. Following treatment, the size of these 2D nanomaterials would be reduced even more, creating quantum dots with unique optical characteristics that might be used in a range of industries. Because of their simplicity of manufacture and outstanding biocompatibility, molybdenum disulfide quantum dots (MoS2 QDs) have gotten a lot of interest among these transition-metal dichalcogenide quantum dots. Furthermore, when compared to other morphologies of MoS2 nanostructures, such as nanosheets and nanotubes, MoS2 QDs demonstrated tiny size and controllable fluorescence emission, endowing them with prospective applications in biology such as biosensing, bioimaging, and so on. Various methodologies, including top-down and bottom-up approaches, have been developed to date for the synthesis of MoS2 QDs. Meanwhile, several applications of MoS2 QDs have been investigated in a variety of fields such as sensing, electrocatalysis, bioimaging, energy, and etc \cite{Arul16}.

To bring our findings in context, we refer to relevant research provided in references \cite{Heinl13,Belouad18}. In the presence of the flux $\phi$, we investigate the propagation of electrons in a circular QD defined electrostatically in  monolayer molybdenium disulfide $\mathrm{MoS_2}$.
Depending on the radius, potential, energy gap, incident energy, and $\phi$, we could identify diﬀerent scattering regimes. We determine the scattering coefficients and the far-field radial component of the reflected current associated with the reflected wave using the boundary condition. 
Under the  suitable conditions, these values exhibit a variety of oscillatory behaviours and resonances, as well as sharp picks at resonance points.

The paper is organized as follows. In section \ref{model}, we set a theoretical model that allows us to describe the wave plane propagation in a circular QD of monolayer molybdenium disulfide $\mathrm{MoS_2}$ subject to a magnetic flux. The solutions of the energy spectrum resulted from Dirac equation are determined by considering two regions and the scattering coefficients are obtained via the boundary condition. In section \ref{cross}, we calculate the scattering efficiency and the far-field radial component of the reflected current in terms of the magnetic flux. Section \ref{Results} will be devoted to the discussions of our numerical results. We conclude our work in the final section.
%========================================================
\section{Band structures}\label{model}
%========================================================
\begin{figure}[!htb]\centering
	\includegraphics[scale=0.3]{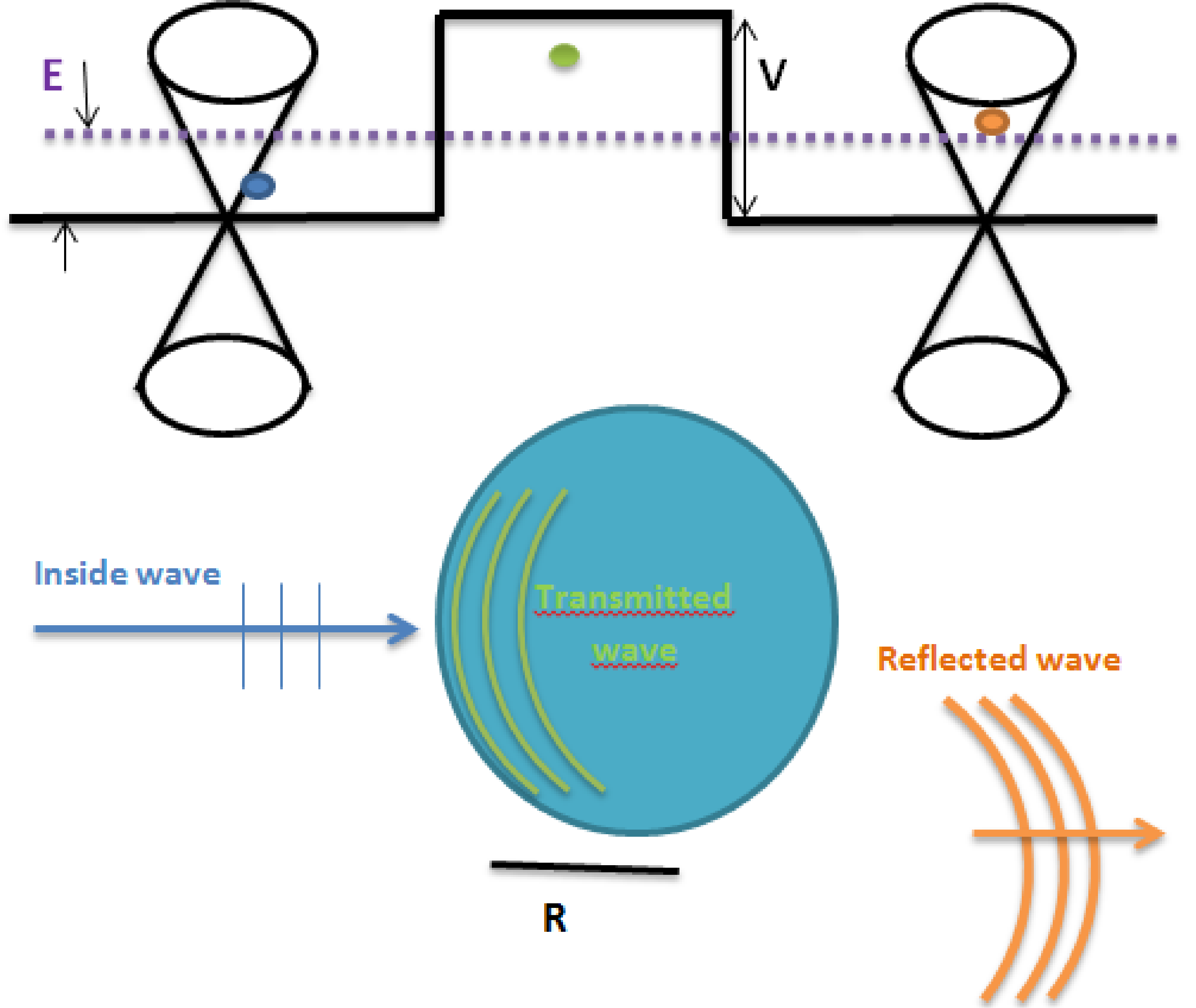}
	\caption{The energy $E$ of the Dirac electrons propagating in monolayer molybdenium disulfide $\mathrm{MoS_2}$ in a circular quantum dot. The dot is characterised by its radius $R$ and the applied bias $V$. The incident plane wave with energy $E>V$ (blue) corresponds to a state in the conduction band (upper cone). The reflected wave (purple) also lies in the conduction band, while the transmitted wave inside the dot (red) occupies the valence band (lower cone).}
	\label{f1}
\end{figure}

We consider a quantum dot of radius $R$ in the presence of a electric potentiel $V$ and  magnetic flux $\phi$, as illustrated in Figure \ref{f1}. In the vicinity of the $k$($\tau=1$) and  $k'$($\tau=-1$) valleys, by constructing the wave functions through the basis of conduction and valence bands, the Dirac-Weyl Hamiltonian for low-energy charge carriers in monolayer molybdenium disulfide $\mathrm{MoS_2}$ reads \cite{Oliveira16, Di Xiao12}
\begin{equation}\label{eq1}
	H=H_0+\frac{\Delta}{2} \sigma_z +\frac{\lambda_{so}}{2}\tau s_z(1-\sigma_z)+V(r)
\end{equation}
so that $H_0$ is equal to
\begin{equation}\label{eq2}
	H_0= -i (\vec \nabla + e\vec A) \cdot \vec \sigma
\end{equation}
where $v_F$ is the Fermi velocity, $\bm{p}\!\!\!=\!\!\!(p_x,p_y)$ is the two-dimensional momentum operator, $\bm{\sigma}\!=\!(\sigma_x,\sigma_y, \sigma_z)$ are Pauli matrices acting on the atomic orbitals, $V(r)$ is the potential barrier, $\Delta=166$ meV \cite{Di Xiao12} is related to the material band gap, $s_z = \pm1$ stands for the electron spin-up  and spin-down, $\tau = \pm 1$ stands for $k$ and $k'$ valleys, $\lambda_{so}=75$ meV \cite{Di Xiao12} is the splitting of the valence band due to spin-orbit coupling and $R$ is the dot radius. Now we'll write  localized-state solutions in our system, which is characterized as a circularly symmetric quantum dot with the potential barrier $V(r)$, energy gap $\Delta(r)$ and vector potential $\vec A(r)$ associated to the magnetic flux $\phi$ measured in quantum unit $\frac{h}{e}$ given by
\begin{equation}
	V(r) =  \left\{
				\begin{array}{cc}
					0, &  r>R \\
					V, &   r\leq R
				\end{array}
			\right., \qquad
	\Delta(r) =  \left\{
					\begin{array}{cc}
						0, &  r>R \\
						\Delta, &  r\leq R
					\end{array}
				\right., \qquad
	\vec A(r)=
	\left\{
	\begin{array}{ll}
		\frac{h}{e}\frac{\phi}{2\pi r}\vec{e}_{\theta},   & r< R \\
    0, &  r>R \\
	\end{array}
	\right.
\end{equation}
where $\vec{e}_{\theta}$ is the unit vector for the azimuthal angle. We carry out our work by considering the polar coordinates (r, $\theta$), such that the Hamiltonian \eqref{eq1}  takes the form
\begin{equation}\label{eq4}
	H = \begin{pmatrix}
			V_+ & \pi^- \\
			\pi^+ & V_-+\tau s_z \lambda_{so} \\
		\end{pmatrix}
\end{equation}
where the two potentials and two operators have been defined
\begin{equation}
	V_{\pm}=V\pm\dfrac{\Delta}{2},\qquad \pi^\pm=\hbar v_Fe^{\pm i\theta}\left(-i\,\frac{\partial}{\partial r}\pm\frac{1}{r}\frac{\partial}{\partial\theta}\pm i\frac{\phi}{r}\right).
\end{equation}
The eigenvalue equation is used to determine the energy spectrum.
\begin{equation}
	H\,\psi_m(r,\theta)=E\,\psi_m(r,\theta).
\end{equation}
Moreover, the commutation between $H$  and total angular momentum, $J_z=-i\,\hbar\,\partial_\varphi+\hbar\,\sigma_z/2$,  allows the separability of the eigenspinors
$\psi_m(r,\theta)$ into the radial $\Phi^{\pm}(r)$ and angular $\digamma^{\pm}(\theta)$ parts. Then, we can write \cite{Schnez08,Heinisch13}
\begin{equation}\label{eq5}
	\psi_m(r,\theta) = \Phi_{m}^{+}(r)\,\digamma_{m}^{+}(\theta) + \Phi_{m+1}^{-}(r)\,\digamma_{m+1}^{-}(\theta)
\end{equation}
where the two angular components are
\begin{equation}\label{eq8}
	\digamma_m^+(\theta)=\frac{e^{im\theta}}{\sqrt{2\pi}}\begin{pmatrix}
															1\\
															0\\
														\end{pmatrix},\qquad
	\digamma_{m+1}^-(\theta)=\frac{e^{i(m+1)\theta}}{\sqrt{2\pi}}\begin{pmatrix}
																	0\\
																	1\\
																\end{pmatrix}
\end{equation}
and $m=0,~ \pm 1,~ \pm 2,~ \cdots$, is the total angular quantum number.
To determine the radial part of the eigenspinors we solve the eigenvalue equation
\begin{equation}
	H\,\psi_m(r,\theta)=E\,\psi_m(r,\theta)
\end{equation}
by considering two regions according to Figure \ref{f1} : outside ($r>R$) and inside ($r\leq R$) the quantum dot. Thus, we have an incident wave $\psi_i$ propagation in the $x$-direction, the reflected wave $\psi_r$ is an outgoing wave and a transmitted wave $\psi_t$ inside the dot. To facilitate our calculation, unless stated otherwise, the following dimensionless quantities:
$r\rightarrow \frac{r}{R}$, $E\rightarrow \frac{E R}{\hbar v_F}$, $\lambda_s\rightarrow \frac{\lambda_s R}{\hbar v_F}$, $V\rightarrow \frac{V R}{\hbar v_F}$ and $\Delta\rightarrow \frac{\Delta R}{\hbar v_F}$.

Outside the dot ($r>R$), the radial components $\Phi_{m}^{+}(r)$ and $\Phi_{m+1}^{-}(r)$ satisfy two coupled differential equations
\begin{eqnarray}
	&&-i\,\frac{\partial}{\partial r}\Phi_{m}^{+}(r)+i\,\frac{m}{r}\,\Phi_{m}^{+}(r)=E\,\Phi_{m+1}^{-}(r)\label{eq10}\\
	&& -i\,\frac{\partial}{\partial r}\Phi_{m+1}^{-}(r)-i\,\frac{m+1}{r}\,\Phi_{m+1}^{-}(r)=E\,\Phi_{m}^{+}(r).\label{eq11}
\end{eqnarray}
which can be handled by injecting \eqref{eq10} into \eqref{eq11} to derive a second differential equation satisfied by $\Phi_{m}^{+}(r)$
\begin{equation}\label{eq12}
	\left(r^2\frac{\partial^2}{\partial^2 r}+r\frac{\partial}{\partial r}+r^2E^2-m^2\right)\Phi_{m}^{+}(r)=0
\end{equation}
showing that the solutions are Bessel functions $J_{m}(Er)$ of type. Moreover, the wave function of the incident electron, propagating along x-direction ($x=rcos\theta$), takes the form
\begin{eqnarray}\label{eq13}
	\psi^i_m(r,\theta) &=& \frac{e^{ikx}}{\sqrt{2}}\left(
               \begin{array}{c}
                 1 \\
                 1 \\
               \end{array}
             \right)=\frac{1}{\sqrt{2}}\sum_m i^m\,
		J_{m}(kr)\,e^{im\theta}
		\begin{pmatrix}
			1\\
			1\\
		\end{pmatrix}
\end{eqnarray}

Using this together with the eigenstates \eqref{eq8}, we write the incident spinor corresponding to the present system as
\begin{eqnarray}\label{eq11bis}
	\psi^i_m(r,\theta) &=& \frac{1}{\sqrt{2}}\sum_m i^m\,
	\left[
		J_{m}(kr)\,e^{im\theta}
		\begin{pmatrix}
			1\\
			0\\
		\end{pmatrix}
		+i\,J_{m+1}(kr)\,e^{i(m+1)\theta}
		\begin{pmatrix}
			0\\
			1\\
		\end{pmatrix}
	\right]
\end{eqnarray}
as well as the reflected wave
\begin{equation}\label{eq12}
	\psi^r_m(r,\theta) = \frac{1}{\sqrt{2}}\sum_m i^m\,\alpha_{m}
	\left[
		H^{(1)}_{m}(kr)\,e^{im\theta}\begin{pmatrix}
												1\\
												0\\
											\end{pmatrix}
		+i\,H^{(1)}_{m+1}(kr)\,e^{i(m+1)\theta}\begin{pmatrix}
												0\\
												1\\
											\end{pmatrix}
	\right]
\end{equation}
wehere  $H^{(1)}_{m}(kr)$ is the Hankel functions of the first kind \cite{Berry13} and $\alpha_{m}$ is the scattering coefficients and the wave number $k=E$.

Inside the dot ($r\leq R$), we obtain the following equations corresponding to the radial functions $\Phi_{m}^{+}$ and $\Phi_{m+1}^{-}$
\begin{eqnarray}
	&& i\left(\frac{\partial}{\partial r}-\frac{m}{r}+\frac{\phi}{r}\right)\Phi_{m}^{+}(r)+\left(E-V_--\tau s_z \lambda_{so}\right)\Phi_{m+1}^{-}(r)=0 \label{eq13}\\
	&& i\left(\frac{\partial}{\partial r}+\frac{m+1}{r}-\frac{\phi}{r}\right)\Phi_{m+1}^{-}(r)+\left(E-V_+ \right)\Phi_{m}^{+}(r)=0 \label{eq14}
\end{eqnarray}
Expressing \eqref{eq13} as
\begin{equation}\label{eq14e}
	\Phi_{m+1}^{-}(r) =  -\frac{i}{E-V_--\tau s_z \lambda_{so}}\left(\frac{\partial}{\partial r}-\frac{m}{r}+\frac{\phi}{r}\right)\Phi_{m}^{+}(r)
\end{equation}
and replacing it in \eqref{eq14} we get a differential equation for $\Phi_{m}^{+}(r)$
\begin{equation}\label{eq15}
	\left(r^2\frac{\partial^2}{\partial^2 r}+r\frac{\partial}{\partial r}+r^2\gamma^2-(m+\phi)^2 \right)\Phi_{m}^{+}(r)=0
\end{equation}
where
\begin{equation}
	\gamma^2=\left(E-V_+\right)(E-V_--\tau s_z \lambda_{so}).
\end{equation}
The solution of \eqref{eq15} can be worked out to get the transmitted wave as
\begin{equation}\label{eq16}
	\psi^t_m(r,\theta) = \frac{1}{\sqrt{2}}\,\sum_m i^m\,\beta_{m}
	\left[
		J_{m+\phi}(\gamma r)\,e^{i(m+\phi)\theta}
		\begin{pmatrix}
			1\\
			0\\
		\end{pmatrix}
		+i\,\mu\,J_{m+\phi+1}(\gamma r)\,e^{i(m+\phi+1)\theta}
		\begin{pmatrix}
			0\\
			1\\
		\end{pmatrix}
	\right]
\end{equation}
where the $\beta_m$ denote the transmission coefficients and
\begin{equation}
	\mu= \sqrt{\frac{E-V_+}{E-V_--\tau s_z \lambda_{so}}}.
\end{equation}
Requiring the eigenspinors continuity at the boundary $r=R$ of the quantum dot,
\begin{equation}
	\psi^i_m(R)+\psi^r_m(R)=\psi^t_m(R),
\end{equation}
to obtain the conditions
\begin{eqnarray}
	&& J_{m}(kR)+\alpha_{m}\,H^{(1)}(kR)=\beta_{m}\,J_{m+\phi}(\gamma R)e^{i\phi\theta }, \label{eq17}\\
	&& J_{m+1}(kR)+\alpha_{m}\,H_{m+1}^{(1)}(kR)=\mu\, \beta_{m}\,J_{m+\phi+1}(\gamma R)e^{i\phi\theta }. \label{eq18}
\end{eqnarray}
Solving these equations to get the scattering coefficients
\begin {equation}\label{eq19}
	\alpha_{m}=-\frac{J_{m+\phi}(\gamma R)\,J_{m+1}(kR)-\mu\, J_{m+\phi+1}(\gamma R)\, J_{m}(kR)}{J_{m+\phi}(\gamma R)\,H^{(1)}_{m+1}(kR)-\mu\, J_{m+\phi+1}(\gamma R)\, H^{(1)}_{m}(kR)}
\end {equation}
and the transmission coefficients by
\begin {equation}\label{eq20}
	\beta_{m}=\frac{J_{m}(kR)\,H^{(1)}_{m+1}(kR)-J_{m+1}(kR)\,H_{m}^{(1)}(kR)}{J_{m+\phi}(\gamma R)\,H^{(1)}_{m+1}(kR)-\mu\, J_{m+\phi+1}(\gamma R)\, H^{(1)}_{m}(kR)}
\end {equation}
\section{Cross section and current density}\label{cross}
Using the scattering coefficients $\alpha_m$ and $\beta_m$ calculated previously, we will illustrate how the current density and cross section associated with the present system are altered by the introduction of a magnetic flux $\phi$. Indeed, the current density is derived from the Hamiltonian \eqref{eq1}.
\begin {equation}\label{eq210}
	{\bm j}=\psi^{\dag}{\bm\sigma} \psi
\end {equation}
where inside the dot we have $\psi=\psi_t$, however, outside the dot we have $\psi=\psi_i+\psi_r $.

The far-field radial component of the reflected current, which characterizes angular scattering, reads
\begin{equation}\label{eq29}
	j_{r}= \psi^{\dag}\left(\sigma_{x} \cos\theta+\sigma_{y}\sin\theta\right)\psi=\psi^{\dag}
	\begin{pmatrix}
		0 & e^{-i\theta} \\
		e^{i\theta} & 0 \\
	\end{pmatrix}
	\psi
\end{equation}
By focusing on the reflected wave, from \eqref{eq29} we show that the corresponding radial current can be written as
\begin{eqnarray}\label{eq23e}
	j^{r}_{r}=\frac{1}{2}\sum^{\infty}_{m=0}A_m(kr) \times
	\begin{pmatrix}
		0 & e^{-i\theta} \\
		e^{i\theta} & 0 \\
	\end{pmatrix}
	\times \sum^{\infty}_{m=0}A_m^*(kr)
\end{eqnarray}
where
\begin{equation}\label{eq24e}
	A_{m}(kr) = (-i)^m\left[
								H_m^{(1)*}(kr)
								\begin{pmatrix}
									\alpha_m^*\,e^{-im\theta} \\
									\alpha^*_{-m-1}\,e^{im\theta}
								\end{pmatrix}
								-i\, H_{m+1}^{(1)*}(kr)
								\begin{pmatrix}
									\alpha^*_{-m-1}\,e^{i(m+1)\theta} \\
									\alpha_m^*\, e^{-(m+1)\theta}
								\end{pmatrix}
							\right]
\end{equation}
the current density \eqref{eq23e} can be calculated using the condition  $kr\gg 1$. In this case, the asymptotic behavior of the Hankel functions is approximated by
\begin{eqnarray}\label{eq26e}
	\displaystyle H_m(kr)\simeq \sqrt{\dfrac{2}{\pi kr}}\,e^{i\left(kr-\frac{m\pi}{2}-\frac{\pi}{4}\right)},
\end{eqnarray}
and it can be injected into \eqref{eq24e}, the current density \eqref{eq23e} can be reduced to the following
\begin{equation}\label{eq24}
	j^{r}_{r}(\theta)=\frac{4}{\pi kr}\sum^{\infty}_{m=0}\left[1+\cos\left(2m+1\right)\theta\right]\,|c_m|^2
\end{equation}
where the coefficients that are involved are
%where
\begin{equation}
	|\zeta_m|^2 = \frac{1}{2}\left[{|\alpha_m|^2 + |\alpha_{-(m+1)}|^2}\right].
\end{equation}
Having derived the radial current for the reflected wave, we now consider another another intresting quantity that is  the scattering cross section. It is given by the total reflected flux through a concentric circle $I_r^r$ divided by the incident flux per unit area $I^i/A_u$ \cite{Grujic11}
\begin{equation}\label{eq28}
\displaystyle\sigma=\dfrac{I_r^r}{I^i/A_u}
\end{equation}
where $I_r^r$ is the total reflected flux through a concentric circle and $(I^i/A_u)$ is the incident flux per unit area. Moreover, $I_r^r$ is given by
\begin{equation}
	I^{r}_{r}=\int_0^{2\pi} j^{r}_{r}(\theta)\,r\,\mathrm{d}\theta=\frac{8}{k}\sum^{\infty}_{m=0}|c_m|^2.
\end{equation}
and we have $I^i/A_u=1$ for the incident wave $\psi_i(x)=\frac{e^{ikx}}{\sqrt{2}}\begin{pmatrix}
1\\
1\\
\end{pmatrix}$. Then it follows that  the scattering cross section coincides with
the total reflected flux.

To give a study of the scattering on the quantum dot with different radius, it is convenient to introduce  the scattering efficiency $Q$. It is defined as the ratio between the scattering cross section $\sigma$ and the geometric one $kR$, giving rise to
 \begin{equation}
Q=\frac{\sigma}{2R}=\frac{4}{kR}\sum^{\infty}_{m=0}|c_m|^2.
\end{equation}
%========================================================
\section{Results and discussions}\label{Results}
\begin{figure}[h!]
	\subfloat[spin-up ($\uparrow$)]{
		\includegraphics[scale=0.62]{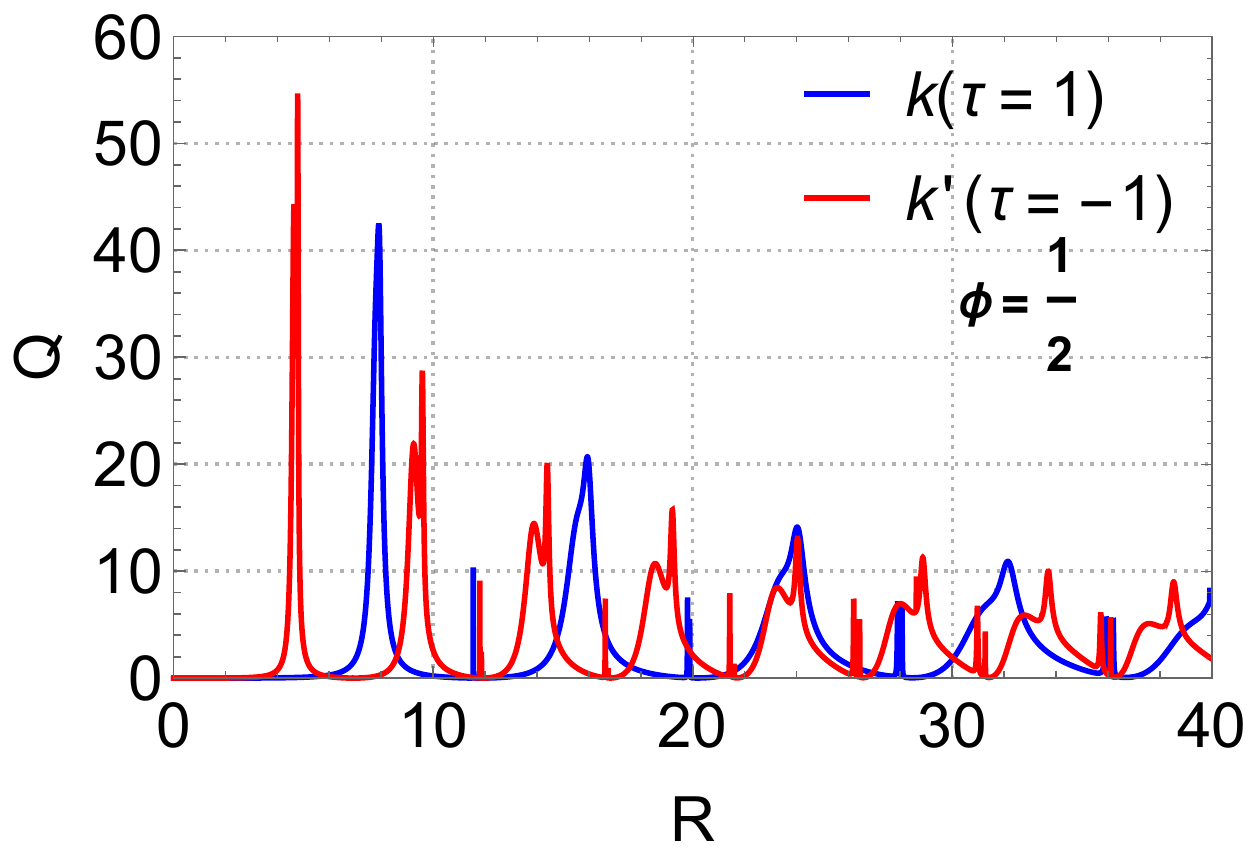}
		\label{Fig:SubFigB}
	}
	\subfloat[spin-down ($\downarrow$)]{
		\includegraphics[scale=0.62]{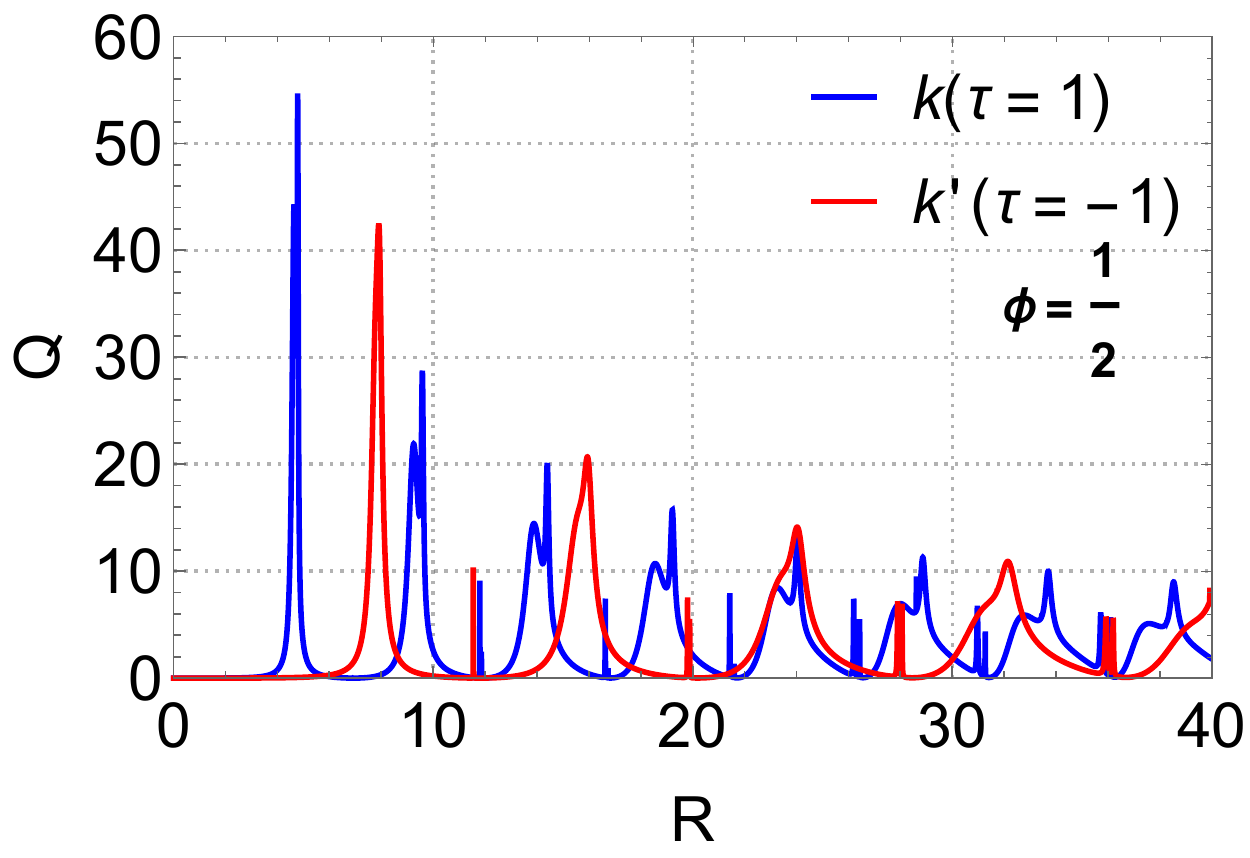}
		\label{Fig:SubFigB}
	}\newline
	\subfloat[spin-up ($\uparrow$)]{
		\includegraphics[scale=0.62]{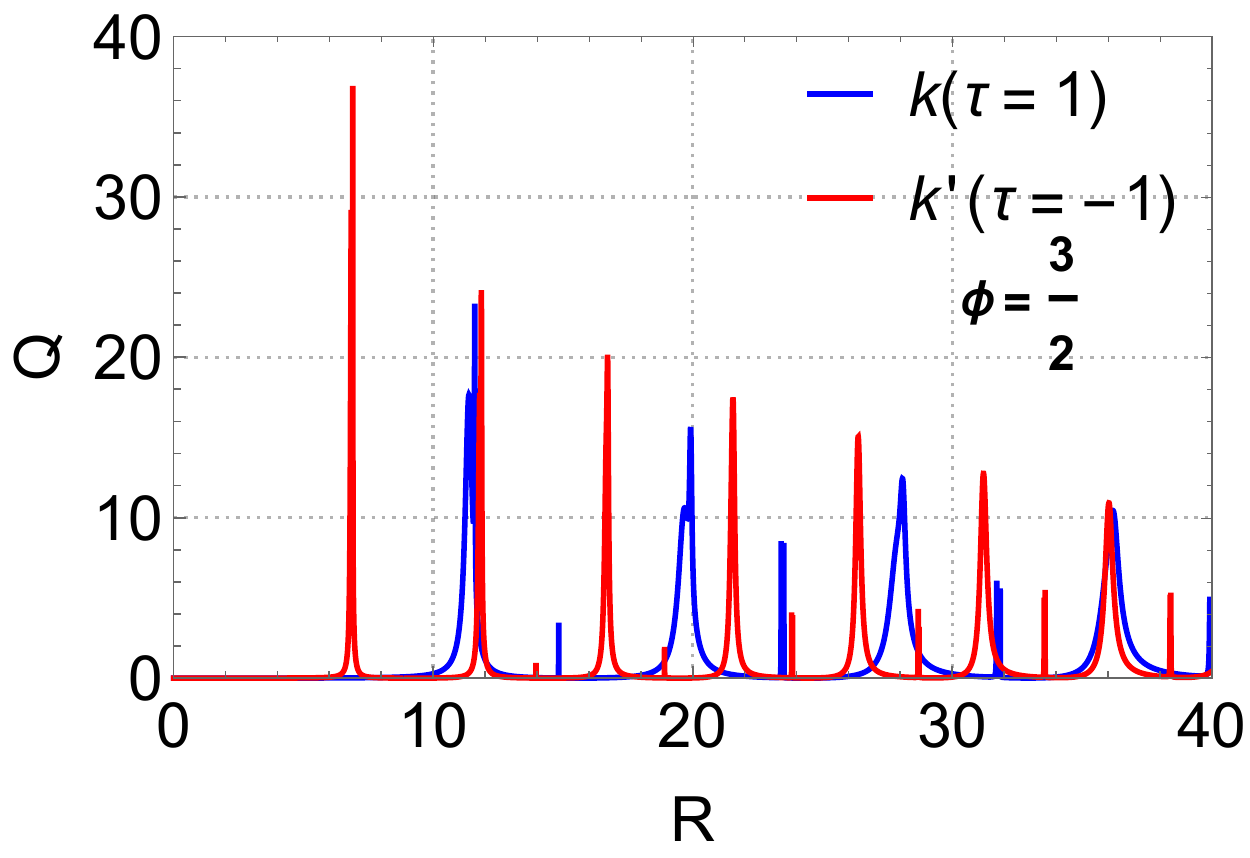}
		\label{Fig:SubFigB}
	}
	\subfloat[spin-down ($\downarrow$)]{
		\includegraphics[scale=0.62]{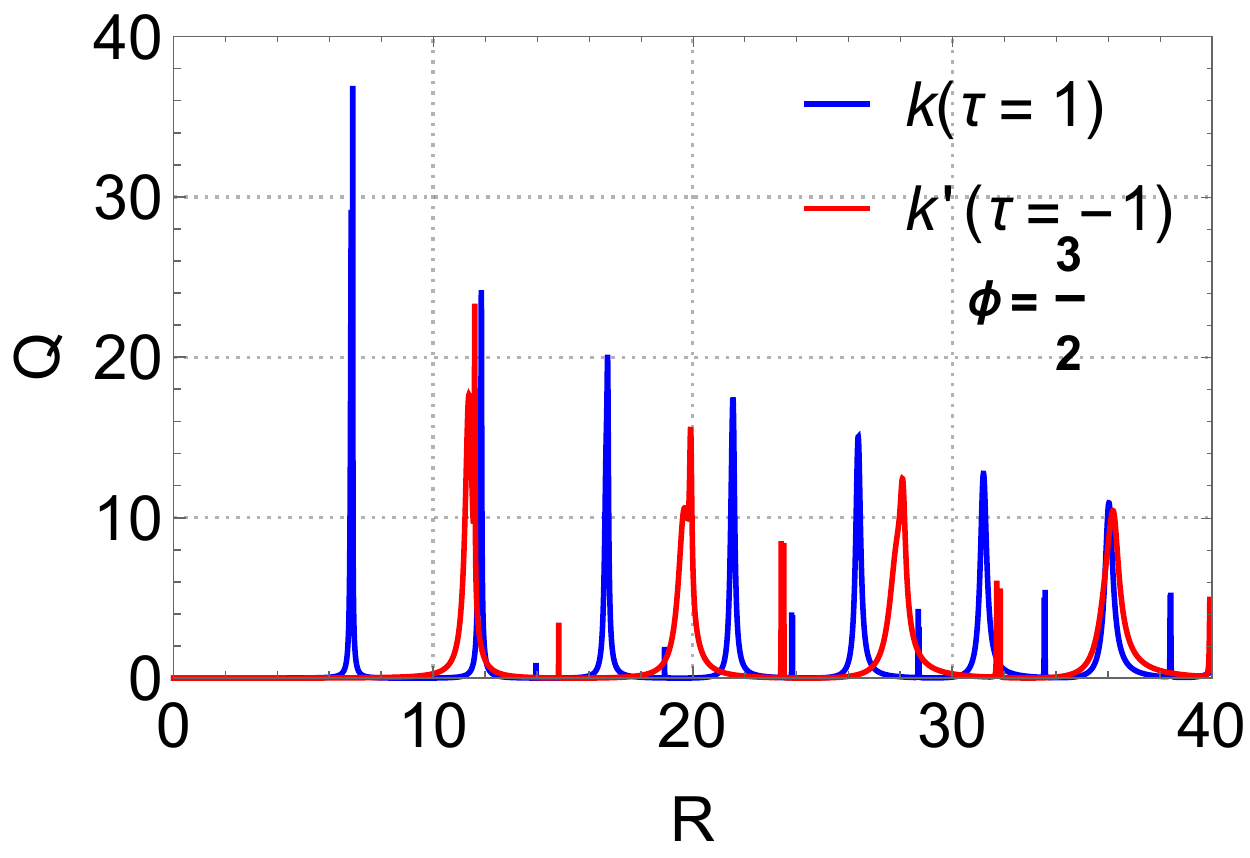}
		\label{Fig:SubFigB}
	}
	\caption{Scattering efficiency $Q$, for the potential $V = 1$, as a function of the dot radius $R$ at $E=0.01$ for (a): the spin-up ($\uparrow$) state  and $\phi=1/2$  in two valleys $k$($\tau=1$) and $k'$($\tau=-1$) and (b): the spin-down ($\downarrow$) state  and $\phi=1/2$  in two valleys $k$($\tau=1$) and $k'$($\tau=-1$) for (c): the spin-up ($\uparrow$)  state  and $\phi=3/2$ in two valleys k($\tau=1$) and $k'$($\tau=-1$) and (d): the spin-down ($\downarrow$)  state  and $\phi=3/2$  in two valleys $k$($\tau=1$) and $k'$($\tau=-1$).}
	\label{Fig2}
\end{figure}

%========================================================
In Figures \ref{Fig2}(a) and \ref{Fig2}(b), we show the scattering efficiency $Q$ as a function of
the radius of the quantum dot $R$, for the quantum states (spin-up ($\uparrow$), $\tau=\pm 1$, $\phi=1/2$) and (spin-down ($\downarrow$), $\tau=\pm 1$, $\phi=1/2$), with $E=0.01$ and $V=1$.
We show that when $R\rightarrow 0$, $Q\rightarrow 0$, when $R$ increases, the scattering efficiency
shows a strongly damped oscillatory behavior and their amplitude decreases significantly with the
 appearance of sharp peaks, similar to graphene quantum dots \cite{Schulz15}.
In figures \ref{Fig2}(c) and \ref{Fig2}(d), we show the scattering efficiency $Q$ as a function of
the radius of the quantum dot $R$, for the quantum states (spin-up ($\uparrow$), $\tau=\pm 1$, $\phi=3/2$) and (spin-down ($\downarrow$), $\tau=\pm 1$, $\phi=3/2$), with $E=0.01$ and $V=1$. The scattering efficiency $Q$ has a sharply damped oscillatory behaviour with the appearance of sharp transverse resonant peaks \cite{Zheng19}, As the radius $R$ increases, the height of the peak decreases but its width becomes larger, which shows that the peculiarity of the energy dispersion becomes apparent. Moreover, by comparing figures \ref{Fig2}(a) and \ref{Fig2}(b) with figures \ref{Fig2}(c) and \ref{Fig2}(d), we find that the dependence of $Q$ for the spin-up ($\uparrow$) and spin-down ($\downarrow$) states in the two valleys $\pm \tau$ depends sharply on $\phi$, as a consequence, the resonance peak height decreases when $\phi$ increases and the resonance peak width increases when $\phi$ increases vice versa. Moreover, we show behaviors of symmetry with respect to $\pm \tau$ i.e.
\begin{equation}
	Q(\uparrow), \tau, \phi)=Q((\downarrow), -\tau, \phi).
\end{equation}

\begin{figure}[!htb]\centering
	\subfloat[spin-up ($\uparrow$)]{
		\includegraphics[scale=0.62]{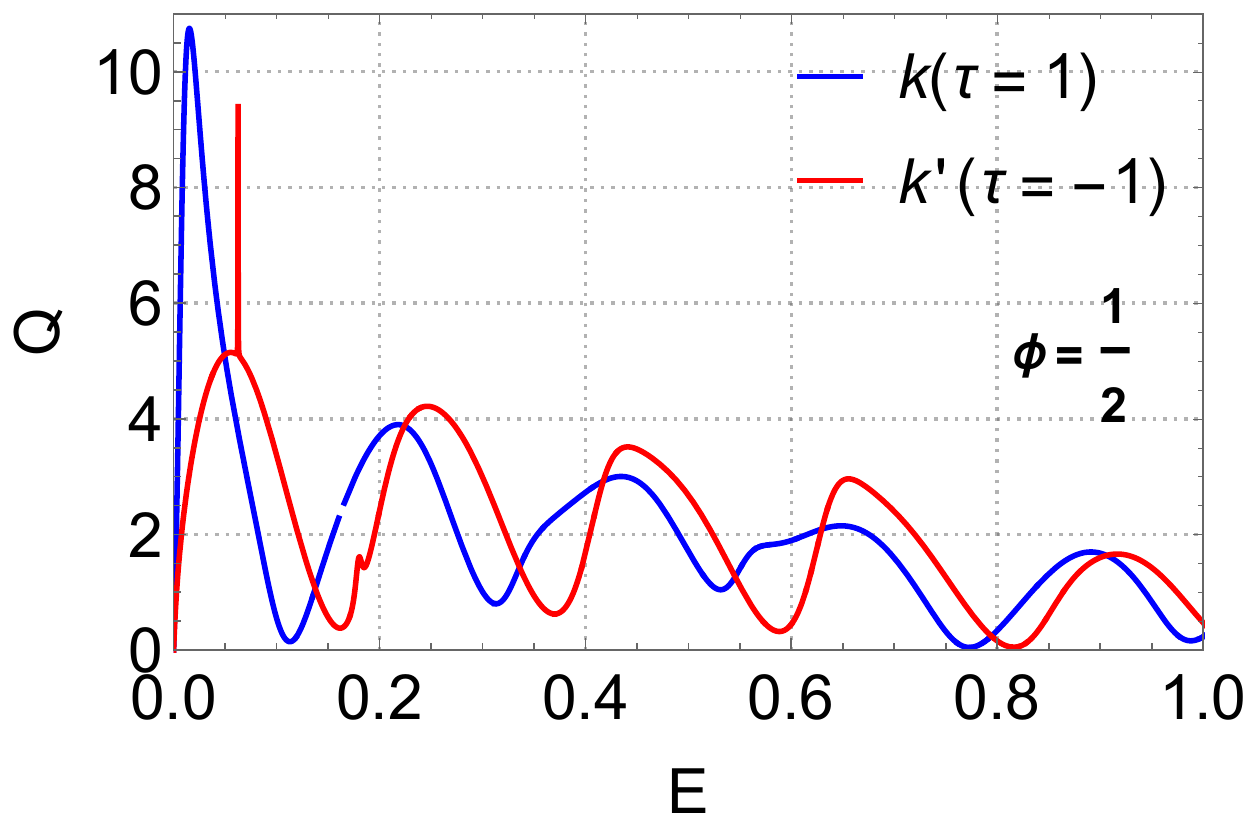}
		\label{Fig:SubFigB}
	}
	\subfloat[spin-down ($\downarrow$)]{
		\includegraphics[scale=0.62]{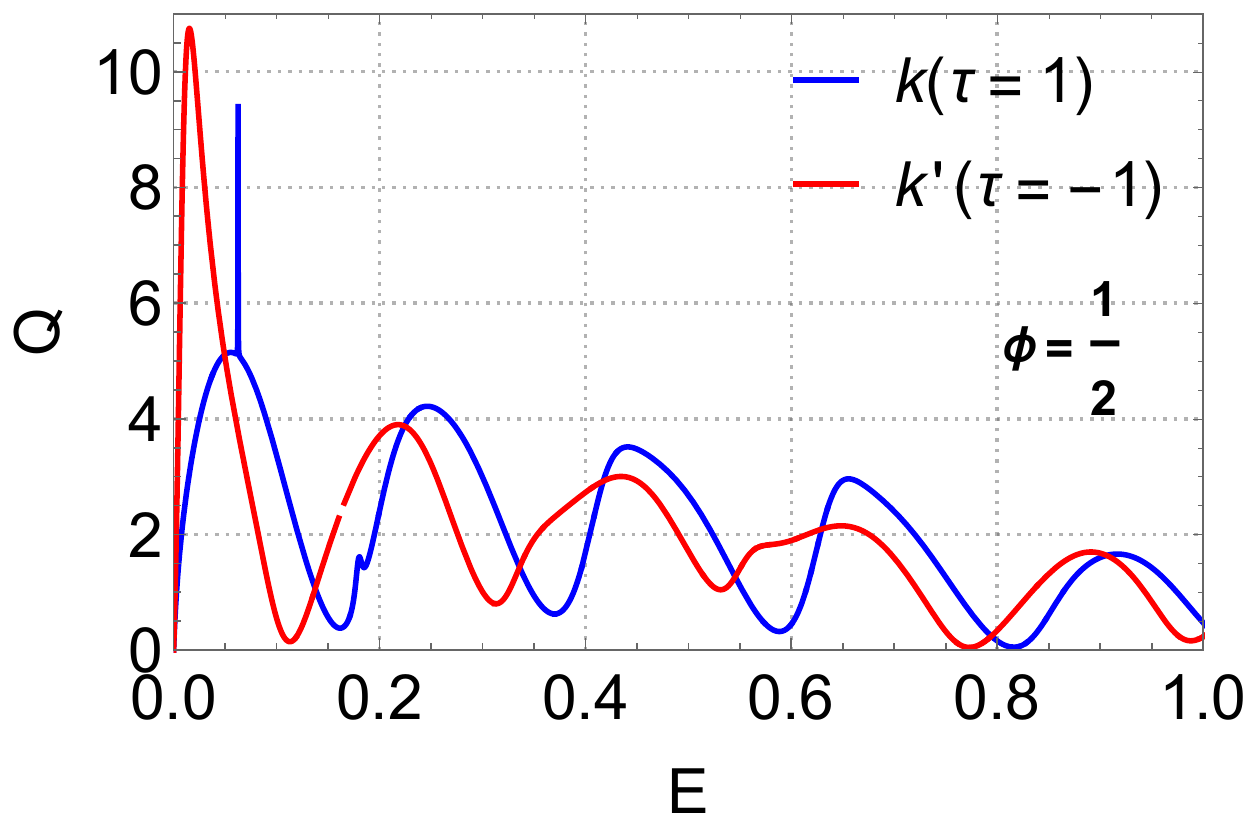}
		\label{Fig:SubFigB}
	}\newline
	\subfloat[spin-up ($\uparrow$)]{
		\includegraphics[scale=0.62]{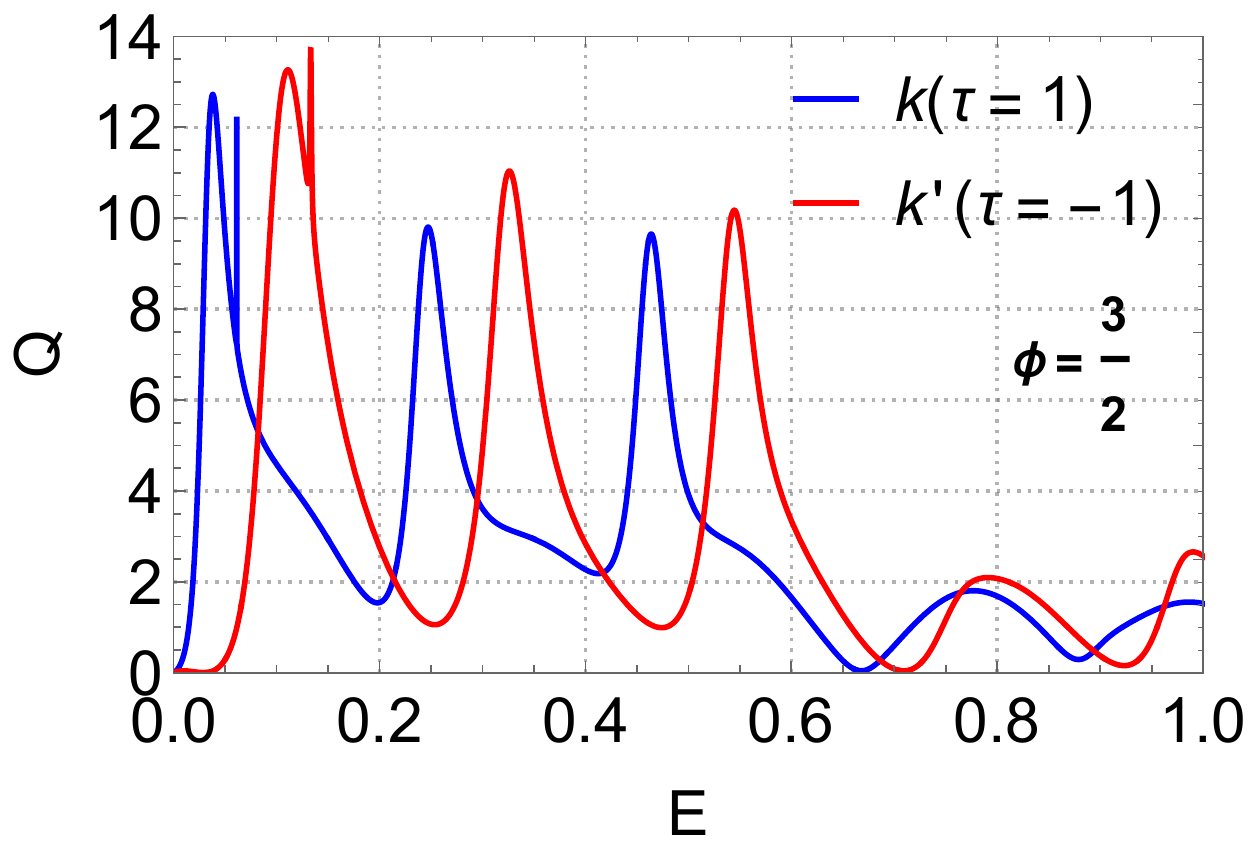}
		\label{Fig:SubFigB}
	}
	\subfloat[spin-down ($\downarrow$)]{
		\includegraphics[scale=0.62]{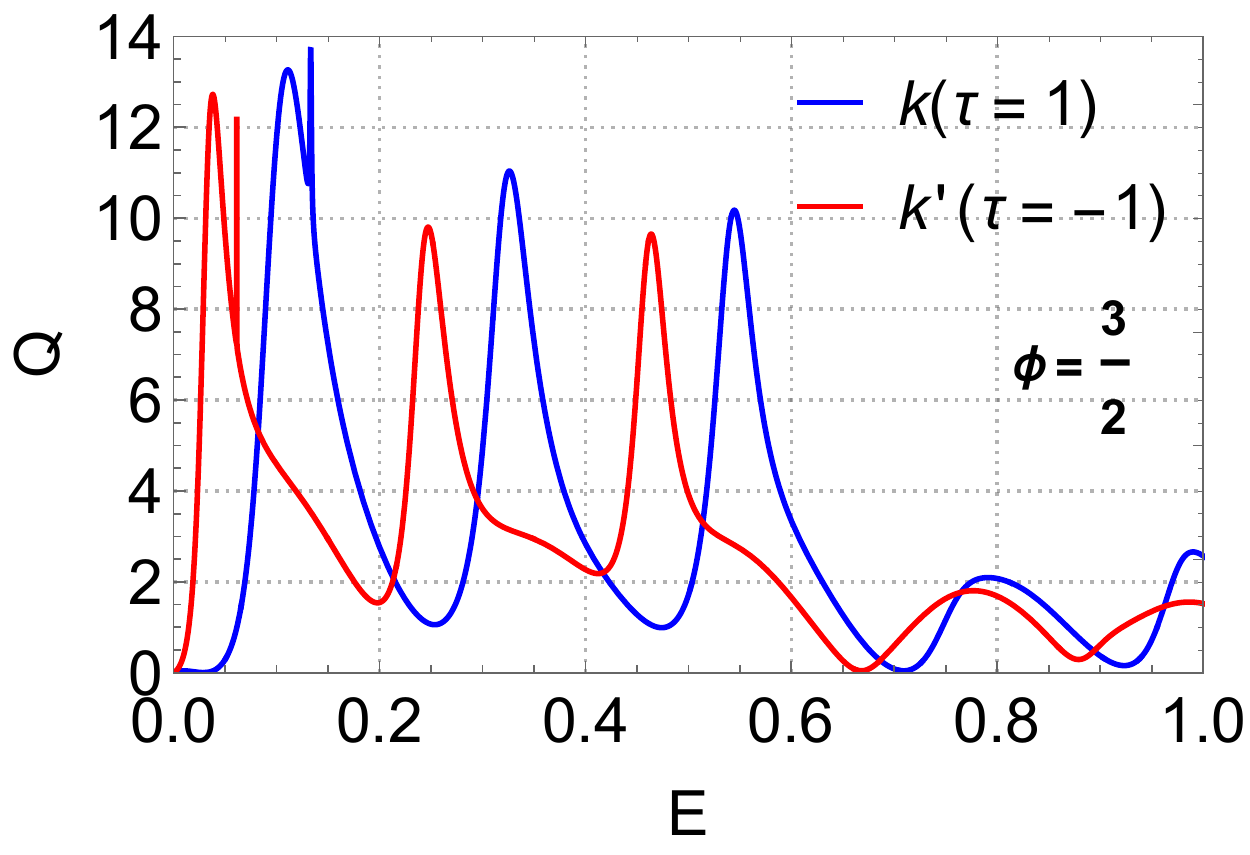}
		\label{Fig:SubFigB}
	}
	\caption{Scattering efficiency $Q$, for the potential $V=1$, as a function of the energy E of the incident electron  at $R=0.9$ for (a):  the spin-up ($\uparrow$) state  and $\phi=1/2$  in two valleys $k$($\tau=1$) and $k'$($\tau=-1$) and (b): the spin-down ($\downarrow$) state  and $\phi=1/2$  in two valleys $k$($\tau=1$) and $k'$($\tau=-1$) for (c): the spin-up ($\uparrow$) state  and $\phi=3/2$  in two valleys $k$($\tau=1$) and $k'$($\tau=-1$) and (d): the spin-down ($\downarrow$) state  and $\phi=3/2$  in two valleys $k$($\tau=1$) and $k'$($\tau=-1$).}
	\label{Fig3}
\end{figure}

Figure \ref{Fig3} shows the scattering efficiency as a function of incident electronic energy $E$ for a quantum dot of radius $R=4$ and applied electric potential $V=1$.
In Figures \ref{Fig3}(a) and \ref{Fig3}(b) we study the states \textbf{($\uparrow \downarrow$,$\tau=\pm 1$,$\phi=1/2$)}: the spin-up ($\uparrow$) and the spin-down ($\downarrow$) for the two valleys $k$($\tau=1$), $k'$($\tau=-1$) with $\phi=1/2$ and Figures \ref{Fig3}(c) and \ref{Fig3}(d) we  consider the states \textbf{($\uparrow \downarrow$, $\tau=\pm 1$, $\phi=3/2$)}: the spin-up ($\uparrow$) and the spin-down ($\downarrow$) for the two valleys $k$($\tau=1$), $k'$($\tau=-1$) with $\phi=3/2$ .

In Figure \ref{Fig3}(a), for $E=0.04$ we show that $Q$ presents a maximum ($Q_{max}=10.75$) for the state ($\uparrow$, $\tau=1$, $\phi=1/2$) with the appearance of a single peak corresponding to $E=0.75$ for the state ($\uparrow$, $\tau=-1$, $\phi=1/2$). For $E>0.05$ we show that $Q$ presents a damped oscillatory behavior \cite{Schulz2015} with the appearance of a single peak adapted to $E=0.175$ for the state ($\uparrow$, $\tau=-1$, $\phi=1/2$) and without appearance of peaks for the state ($\uparrow$,$\tau=1$,$\phi=1/2$). In Figure \ref{Fig3}(b), for $E=0.05$ we show that $Q$ exhibits a maximum for the state ($\downarrow$, $\tau=1$, $\phi=1/2$) with the appearance of a single peak corresponding to $E=0.75$ and without appearance of peaks for the state ($\downarrow$, $\tau=-1$, $\phi=1/2$). For $E>0.05$ we show that $Q$ exhibits damped oscillatory behavior with the appearance of a single peak fitted at $E=0.175$ for the ($\downarrow$, $\tau=1$, $\phi=1/2$) state and without appearance of peaks for the ($\downarrow$,$\tau=-1$, $\phi=1/2$) state. Consequently, the electron scattering efficiency is invariant under the transformation Q($\uparrow$, $\tau$, $\phi$)=Q($\downarrow$, $-\tau$,$\phi$) when $\tau \rightarrow -\tau$ and spin-up ($\uparrow$) tends to spin-down ($\downarrow$).

In Figures \ref{Fig3}(c) and \ref{Fig3}(d), we observe that $Q$ also shows large maxima ($Q_{max}=13.75$) for low energies adapted to $E=0.125$ with the appearance of peaks emerging for the quantum states ($\uparrow$, $\pm\tau$, $\phi=3/2$) and ($\downarrow$, $\pm\tau$, $\phi=3/2$) respectively, these sharp peaks are due to the resonant excitation of the normal modes of the quantum dot, But when E increases, one observes damped oscillations for the spin-up ($\uparrow$) and spin-down ($\downarrow$) states. However, $Q$ is symmetric about $\pm \tau$ i.e. Q($\uparrow$, -$\tau$, $\phi$)= Q($\downarrow$, $\tau$, $\phi$). Consequently the amplitude $Q_{max}(\phi=3/2) < Q_{max}(\phi=1/2)$ and the behavior of $Q$ depends strongly on $\phi$ as shown in Figures \ref{Fig3} (a,b) and Figures \ref{Fig3} (c,d).

\begin{figure}[!htb]\centering
	\subfloat[spin-up ($\uparrow$)]{
		\includegraphics[scale=0.62]{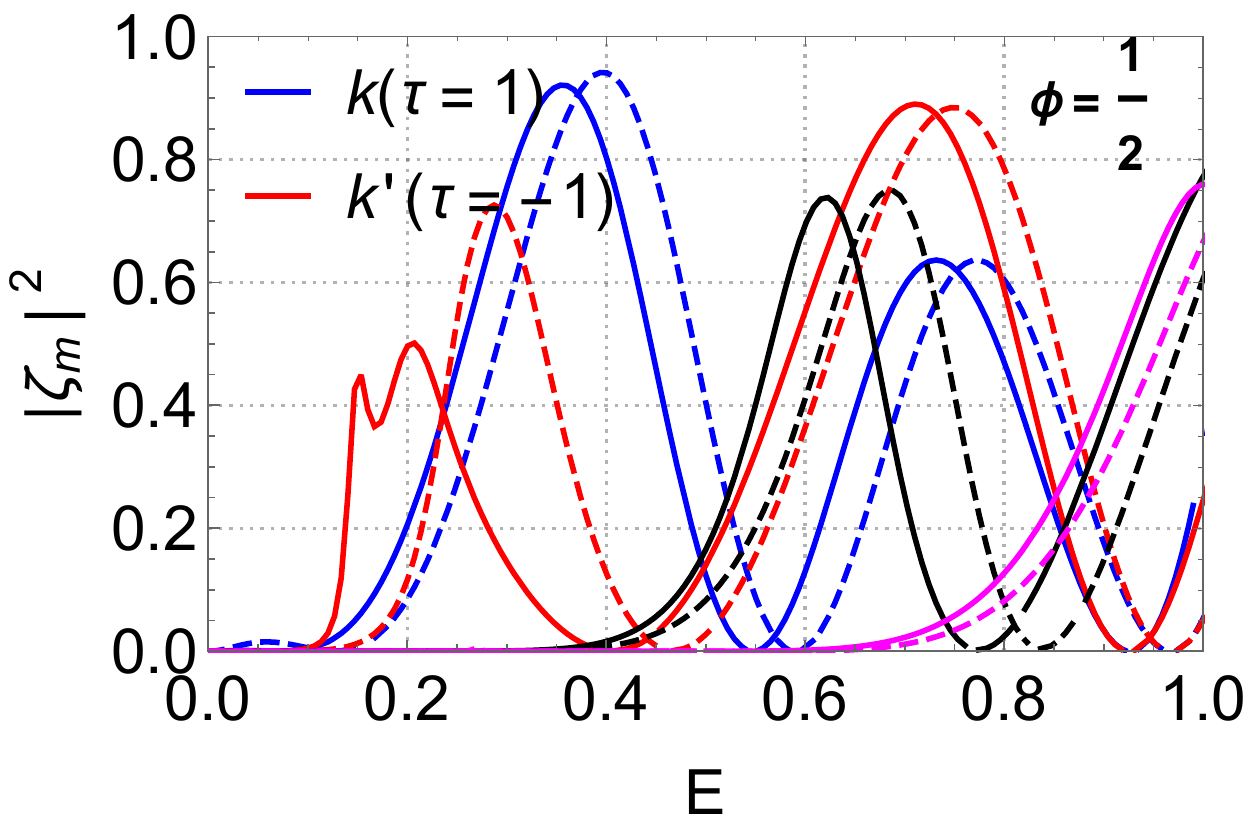}
		\label{Fig:SubFigB}
	}
	\subfloat[spin-down ($\downarrow$)]{
		\includegraphics[scale=0.62]{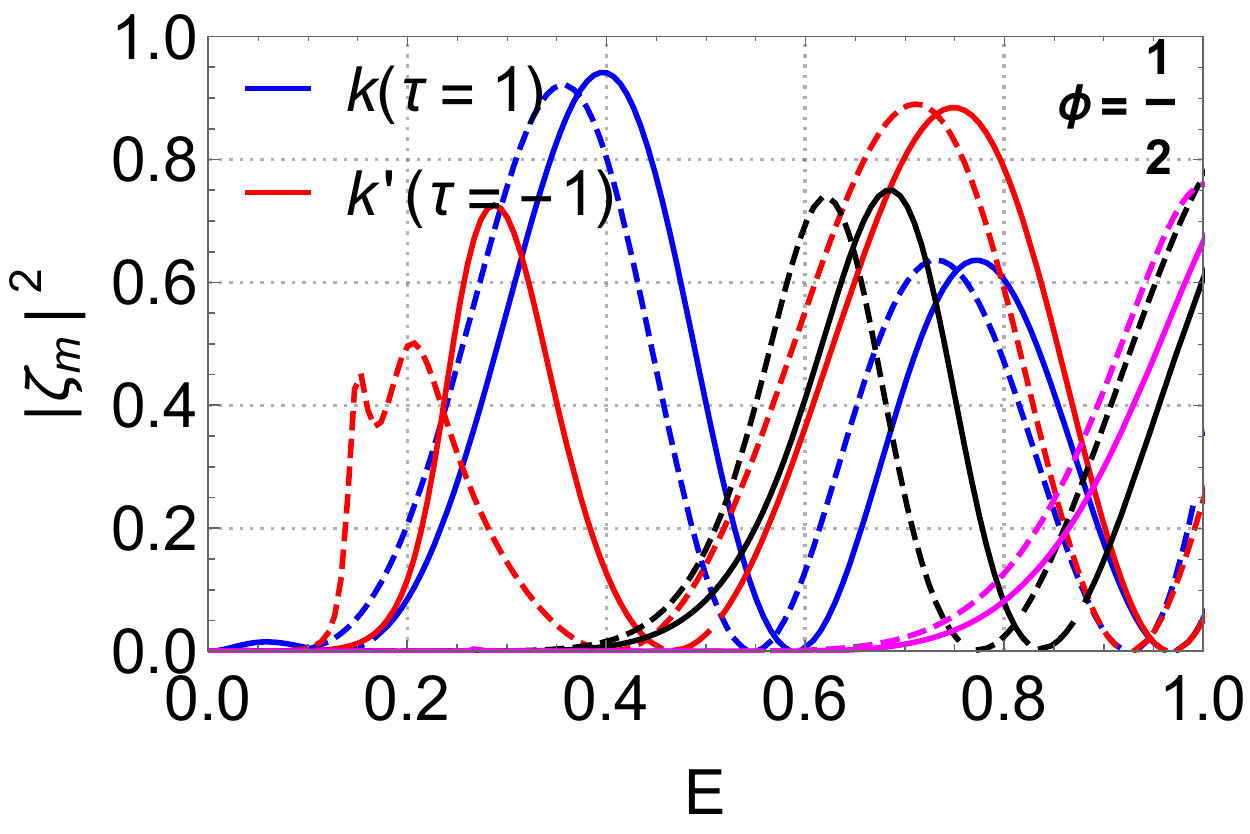}
		\label{Fig:SubFigB}
	}\newline
	\subfloat[spin-up ($\uparrow$)]{
		\includegraphics[scale=0.62]{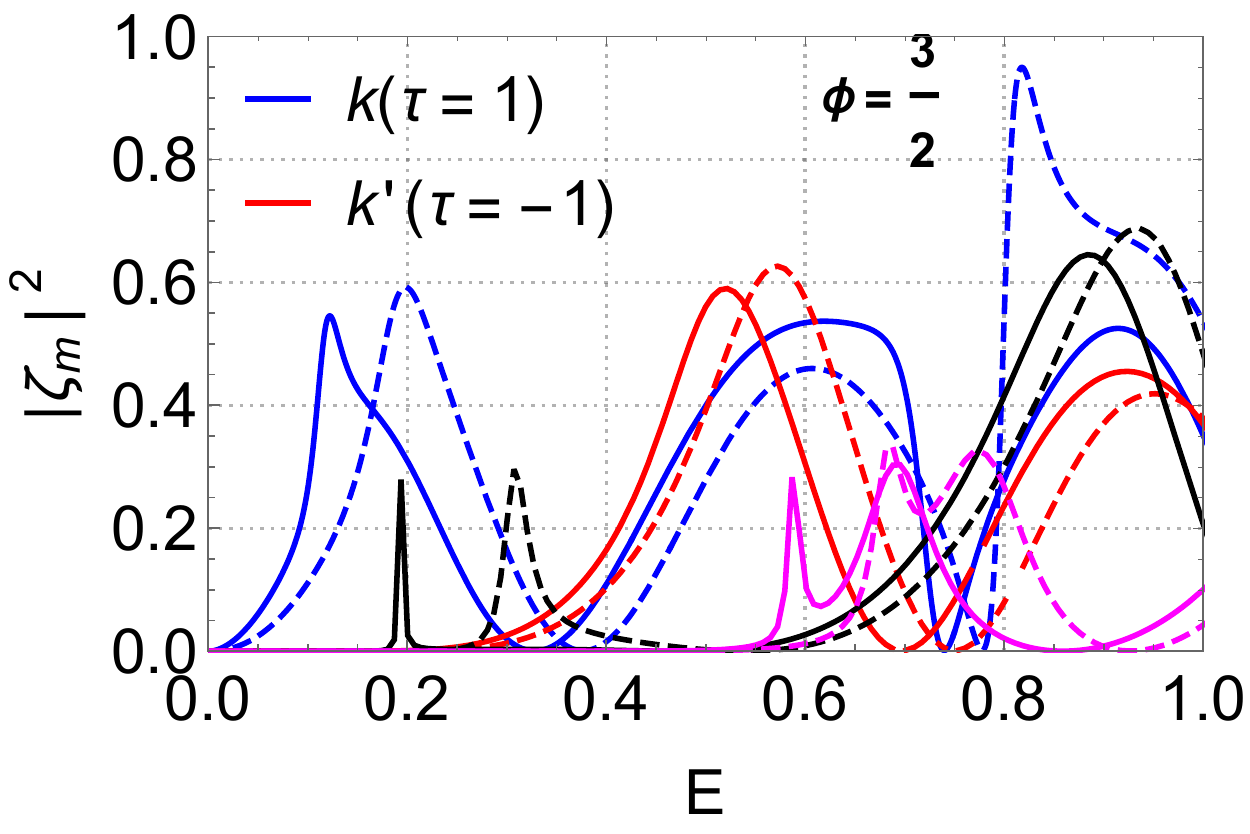}
		\label{Fig:SubFigB}
	}
	\subfloat[spin-down ($\downarrow$)]{
		\includegraphics[scale=0.62]{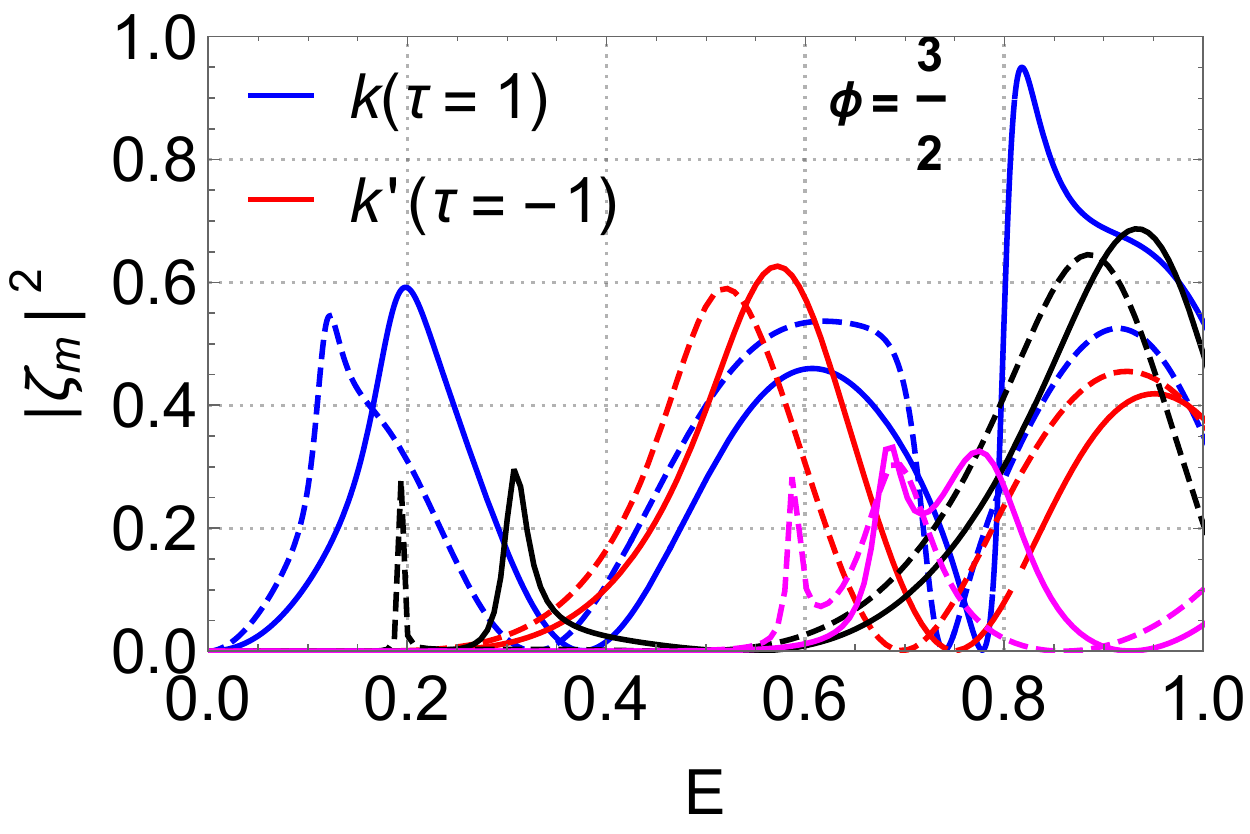}
		\label{Fig:SubFigB}
	}
	\caption{Square modulus of the scattering coefficients $|\zeta_m|^2$, with the potential $V=1$ and $R=2$,  for $m=0$ (blue line), $1$ (red line), $2$ (green line), $3$ (magenta line) as a function of the energy $E$ for (a): ($k$, $k'$, spin-up ($\uparrow$), $\phi=1/2$) states, (b): ($k$, $k'$, spin-down ($\downarrow$), $\phi=1/2$) states, (c): ($k$, $k'$, spin-up ($\uparrow$), $\phi=3/2$) states, (d): ($k$, $k'$, spin-down ($\downarrow$), $\phi=3/2$) states, (e): ($k$, $k'$, spin-up ($\uparrow$)) states and (f): ($k$, $k'$, spin-down ($\downarrow$)) states. Solid (dashed) line corresponds to valley $k$ ($k'$).}
	\label{Fig4}
\end{figure}

To distinguish the resonances, we plot the square modulus of the scattering coefficients $|\zeta_m|^2$ as a function of the energetics in Figure \ref{Fig4}, for near $E=0$, only the lowest scattering coefficient $\zeta_0$ is non-zero, higher order scattering coefficients $\zeta_m$ ($m>0$) as energy increases, as the energy E increases,the $|\zeta_m|^2$ exhibits oscillatory behavior, with consecutive modes appearing interleaved with sharp peaks of distinct $|\zeta_m|^2$ \cite{Schulz15}. The results of Figures \ref{Fig4}(a) and \ref{Fig4}(b) show that the square module of diffusion coefficients $|\zeta_m|^2$ is related to the symetric relation $|\zeta_m|^2$($\uparrow$, $\tau$, $\phi=1/2$)=$|\zeta_m|^2$($\downarrow$,$-\tau$,$\phi=1/2$).

Figures \ref{Fig4}(c) and \ref{Fig4}(d) prove the same results that were noted previously in Figures \ref{Fig4}(a) and \ref{Fig4}(b), with the obvious change with respect to the amplitude of the $|\zeta_m|^2$, that is: $|\zeta_m|^2(\phi=1/2)\neq |\zeta_m|^2(\phi=3/2)$. In a more general way $|\zeta_m|^2$ check the symmetrical: $|\zeta_m|^2$($\uparrow$, $\tau$, $\phi$=$|\zeta_m|^2$($\downarrow$,$-\tau$,$\phi$) and the antisymmetrical: $|\zeta_m|^2(\uparrow, \tau, \phi) \neq |\zeta_m|^2(\downarrow, \tau, \phi)$.
\begin{figure}[H]\centering
	\subfloat[spin-up ($\uparrow$)]{
		\includegraphics[scale=0.62]{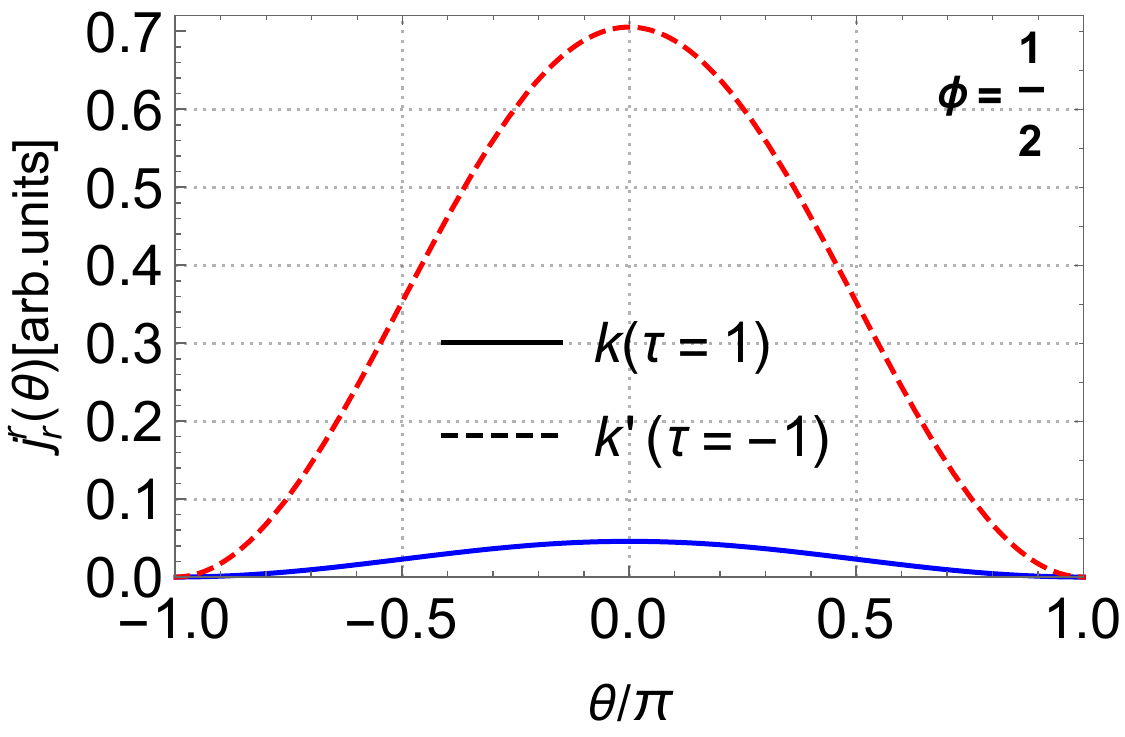}
		\label{Fig:SubFigB}
	}
	\subfloat[spin-down ($\downarrow$)]{
		\includegraphics[scale=0.62]{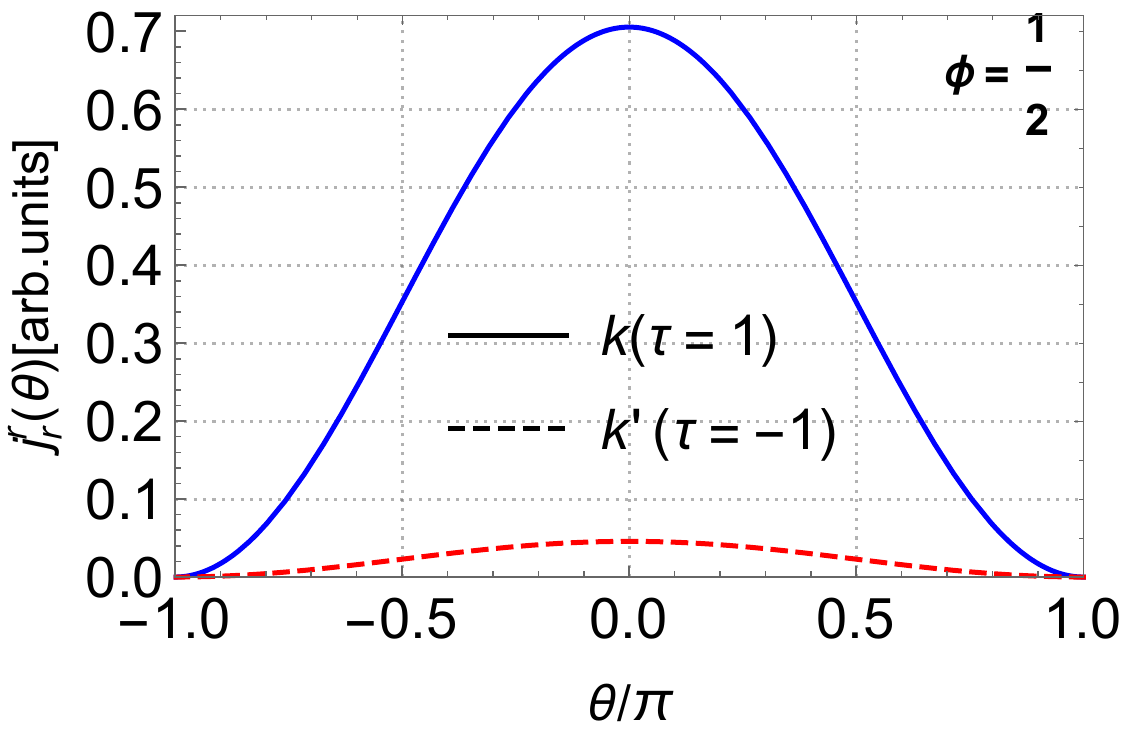}
		\label{Fig:SubFigB}
	}\newline
	\subfloat[spin-up ($\uparrow$)]{
		\includegraphics[scale=0.62]{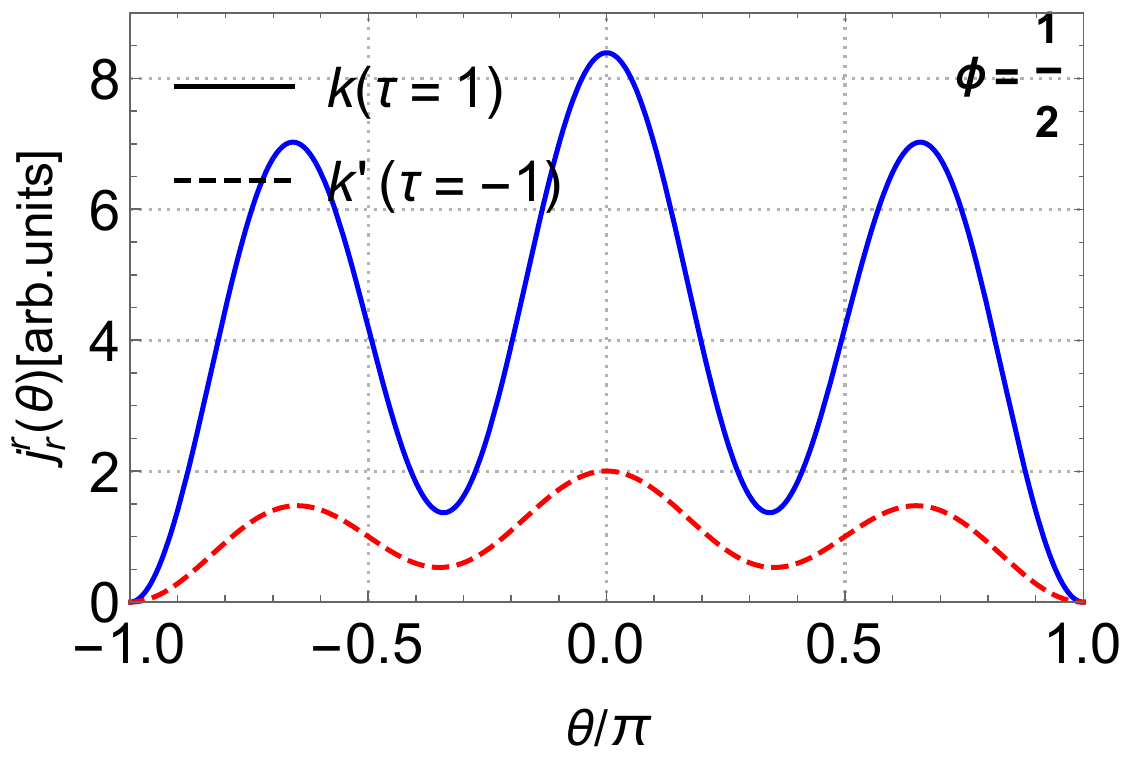}
		\label{Fig:SubFigB}
	}
	\subfloat[spin-down ($\downarrow$)]{
		\includegraphics[scale=0.62]{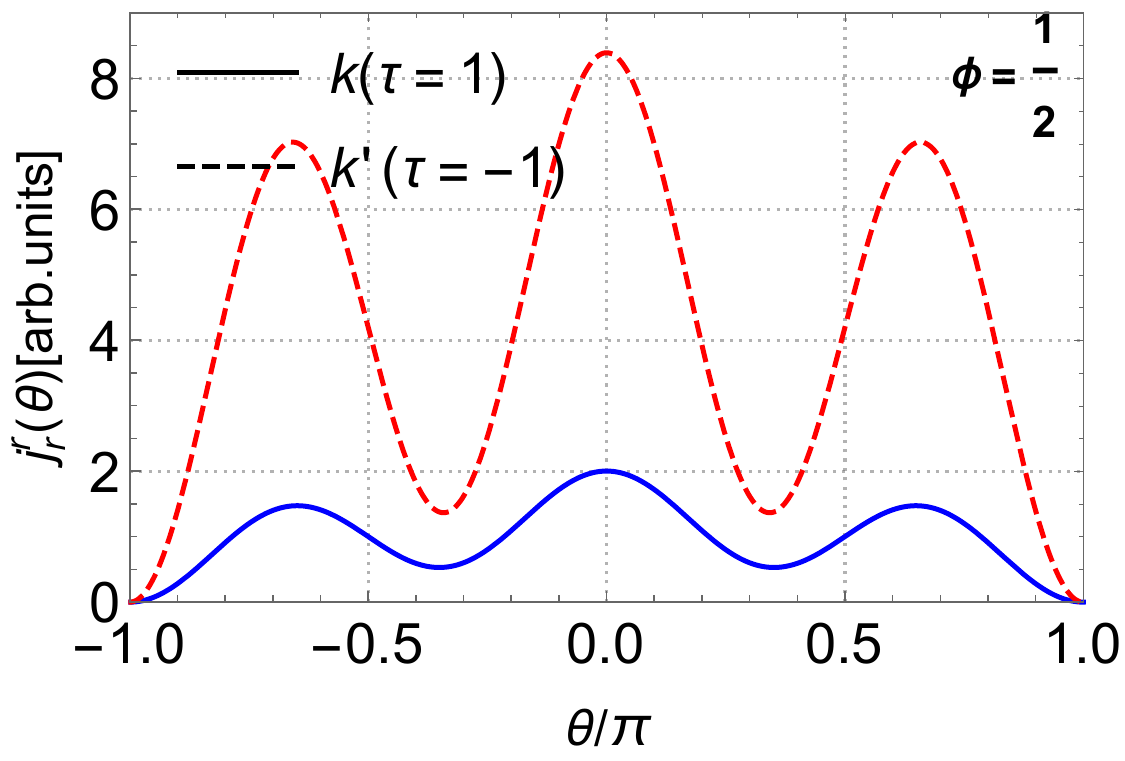}
		\label{Fig:SubFigB}
	}\newline
	\subfloat[spin-up ($\uparrow$)]{
		\includegraphics[scale=0.62]{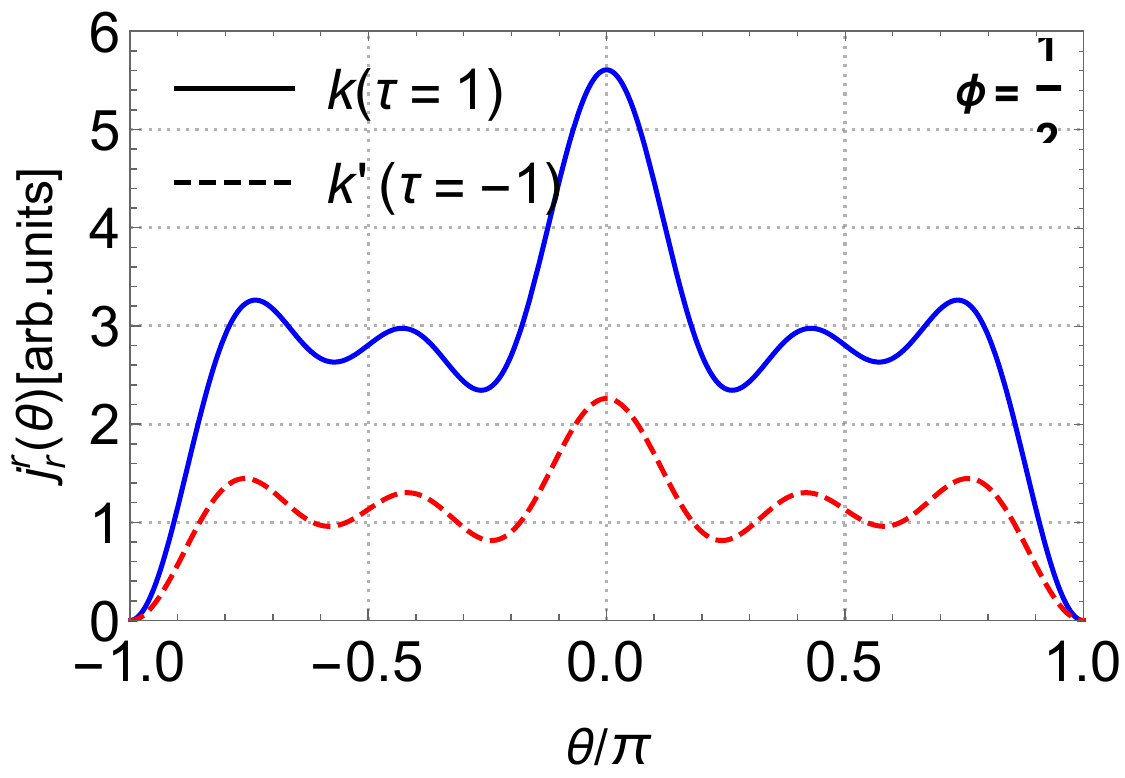}
		\label{Fig:SubFigB}
	}
	\subfloat[spin-down ($\downarrow$)]{
		\includegraphics[scale=0.62]{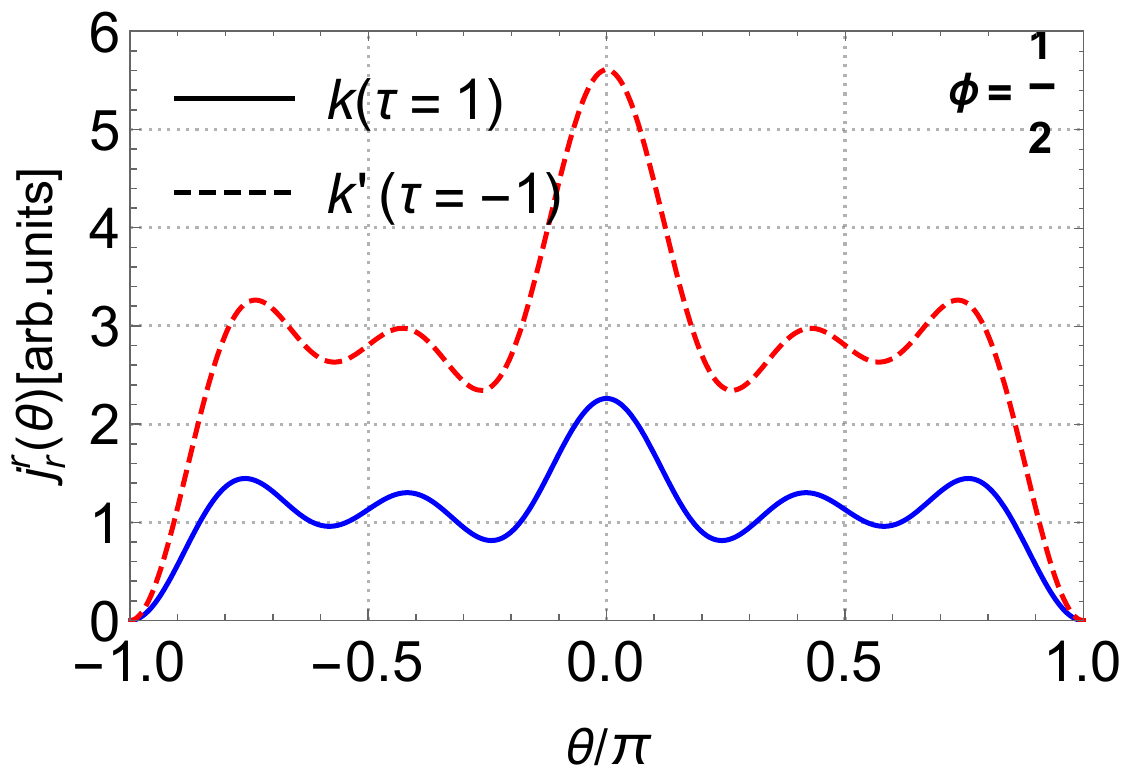}
		\label{Fig:SubFigB}
	}\newline
	\subfloat[spin-up ($\uparrow$)]{
		\includegraphics[scale=0.62]{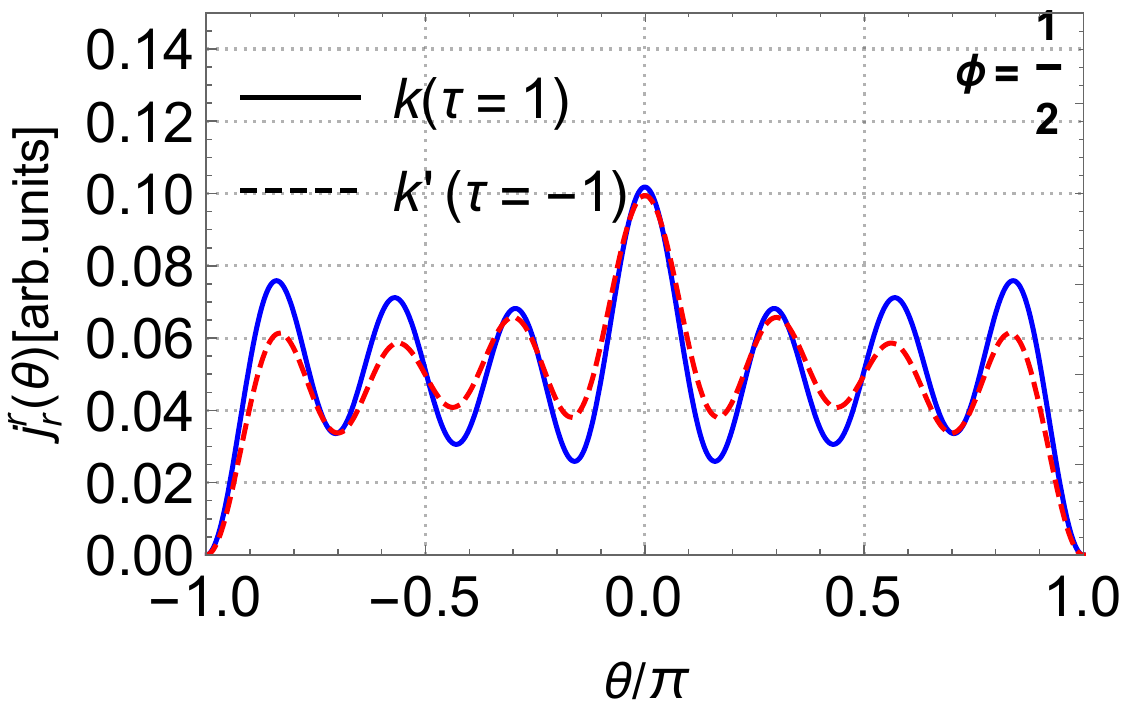}
		\label{Fig:SubFigB}
	}
	\subfloat[spin-down ($\downarrow$)]{
		\includegraphics[scale=0.62]{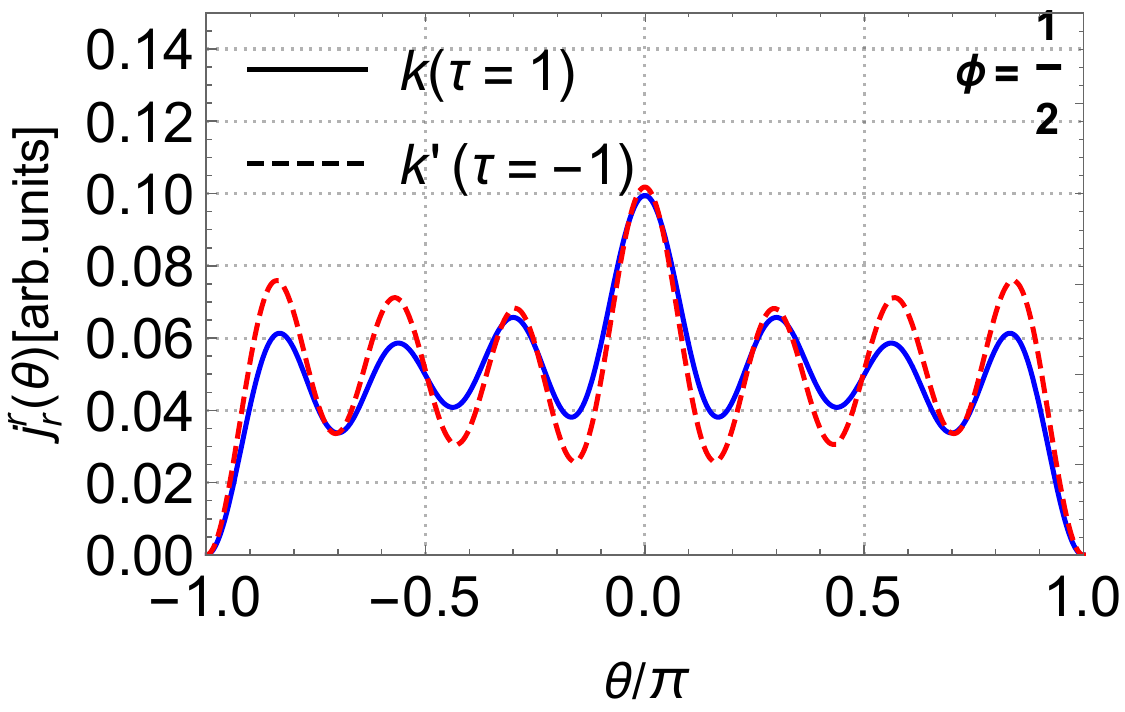}
		\label{Fig:SubFigB}
	}
	\caption{Radial component of the far-field scattered current $j^r_r$ as a function of the angle $\theta$ for (a): ($k$, $k'$, spin-up ($\uparrow$), $m=0$) and (b): ($k$, $k'$, spin-down ($\downarrow$, $m=0$) states, (c): ($k$, $k'$, spin-up ($\uparrow$), $m=1$) and (d): ($k$, $k'$, spin-down ($\downarrow$), $m=1$) states, (e): ($k$, $k'$, spin-up ($\uparrow$), $m=2$)and (f): ($k$, $k'$, spin-down ($\downarrow$), $m=2$) states, (g): ($k$, $k'$, spin-up ($\uparrow$), $m=3$) and (h): ($k$, $k'$, spin-down ($\downarrow$), $m=3$) states with $R=7.75$, $V=1$ , $\phi=1/2$ and $E=0.0704$.}
	\label{Fig5}
\end{figure}

We depict in Figure \ref{Fig5} the angular characteristic of the reflected radial component $j^{r}_{r}$ as a function of $\theta$  for the quantum states ($\uparrow$, $\pm \tau$, $\phi=1/2$) and ($\downarrow$, $\pm \tau$, $\phi=1/2$) with $R=7.25$ and $E=0.0704$ for different values of angular momentum $m$ ($m=0, 1, 2, 3$). We show that $j^{r}_{r}$ exhibits a maximum for $\theta=0$  and a minimum for $\theta=\pm \pi$  for each value of $m$. Moreover, for the mode $m=0$ ( Figure \ref{Fig5}(a) and Figure \ref{Fig5}(b)) only one direction of diffusion towards the front is preferred, while the mode $m=1$, one has three different directions of diffusion, thus the five preferred directions of diffusion of the mode $m=2$ are illustrated on the Figure \ref{Fig5} (c) and Figure \ref{Fig5}(d), moreover the mode $m=3$, has seven preferred directions of diffusion are always observable but with different amplitudes (Figure \ref{Fig5}(e) and Figure \ref{Fig5}(f)). In general, for each mode $m$ we observe $2m+1$ preferred diffusion direction with different amplitudes [40], however the mode ($m=0$) has a greater amplitude than the higher modes ($m>0$). Therefore, in both the spin-up and spin-down states, the dependence of $j^{r}_{r}$ on $\tau$ is symmetric about $\pm\tau $, i.e., $j^{r}_{r}(\uparrow, \tau, \phi)= j^{r}_{r}(\downarrow,  \tau, \phi)$.

\begin{figure}[]\centering
	\subfloat[spin-up ($\uparrow$)]{
		\includegraphics[scale=0.62]{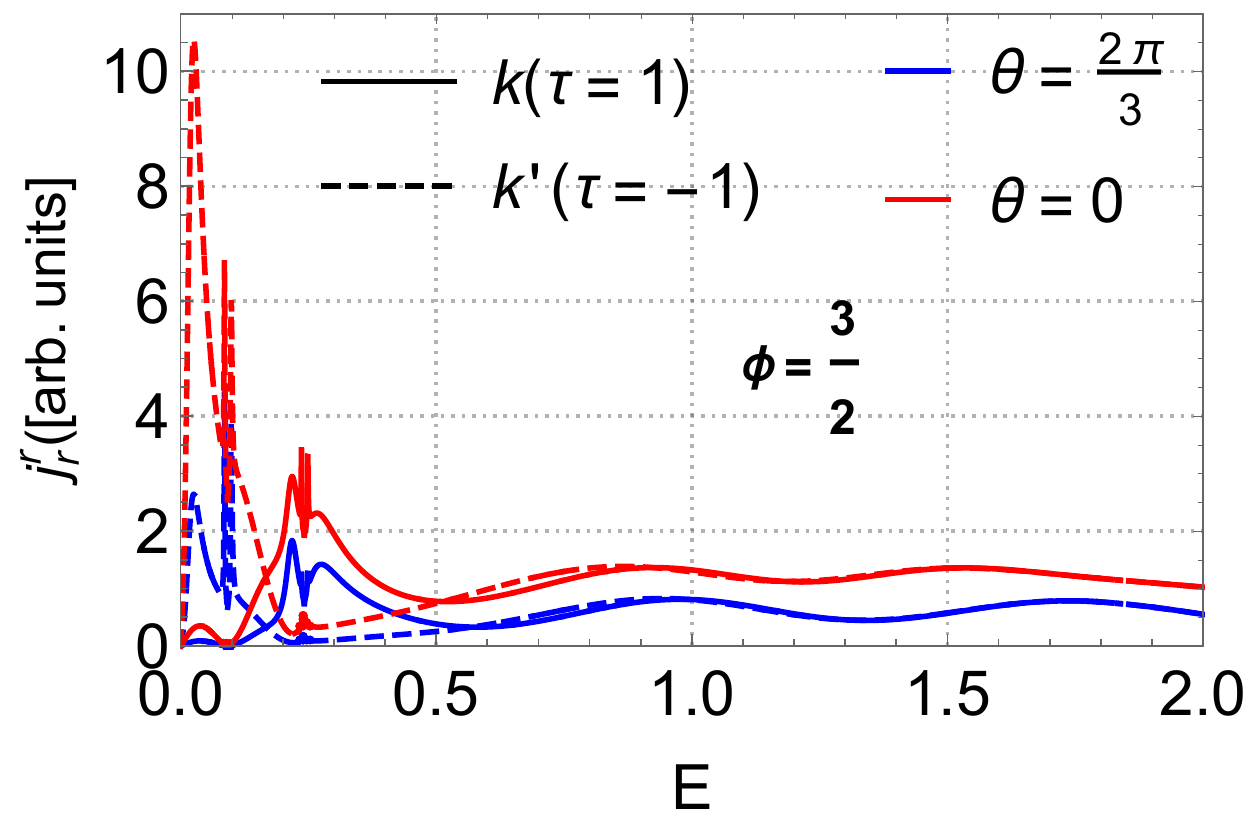}
		\label{Fig:SubFigB}
	}
	\subfloat[spin-down ($\downarrow$)]{
		\includegraphics[scale=0.62]{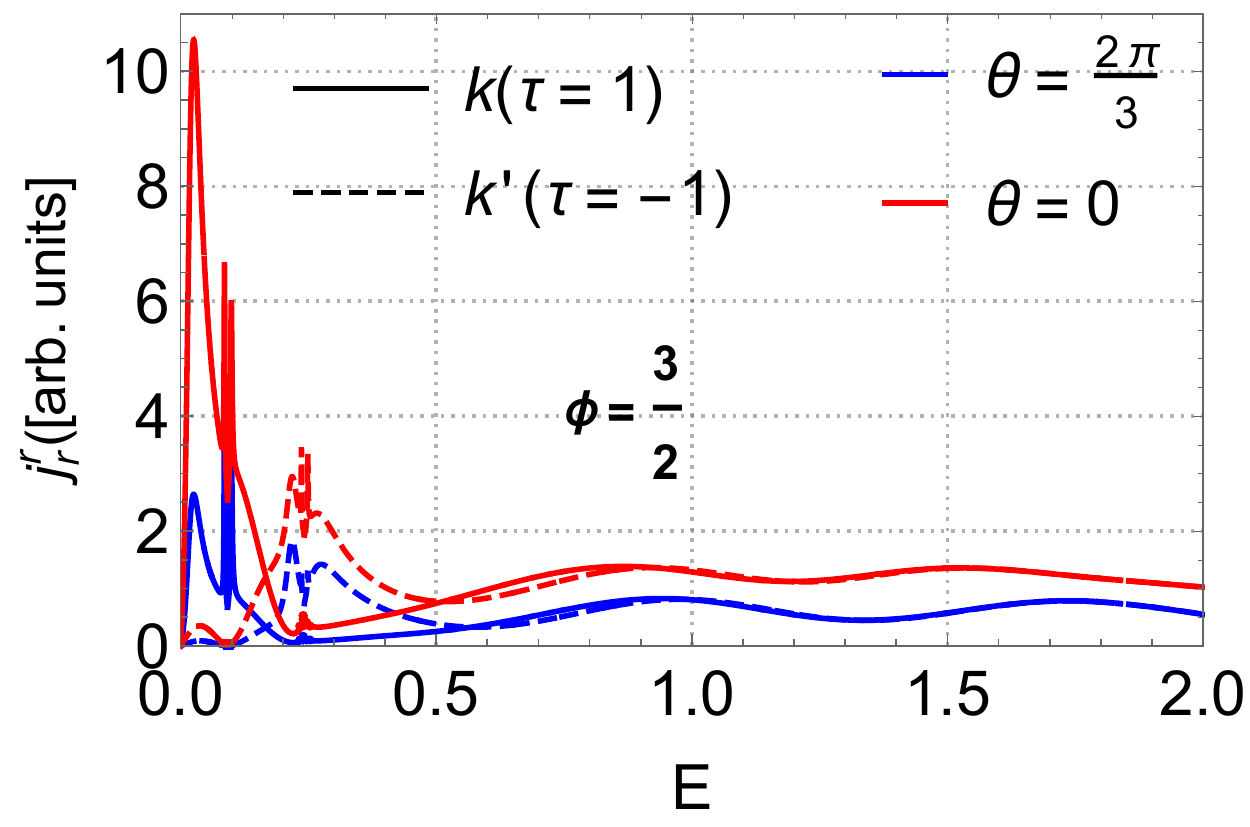}
		\label{Fig:SubFigB}
	}\newline
	\subfloat[spin-up ($\uparrow$)]{
		\includegraphics[scale=0.62]{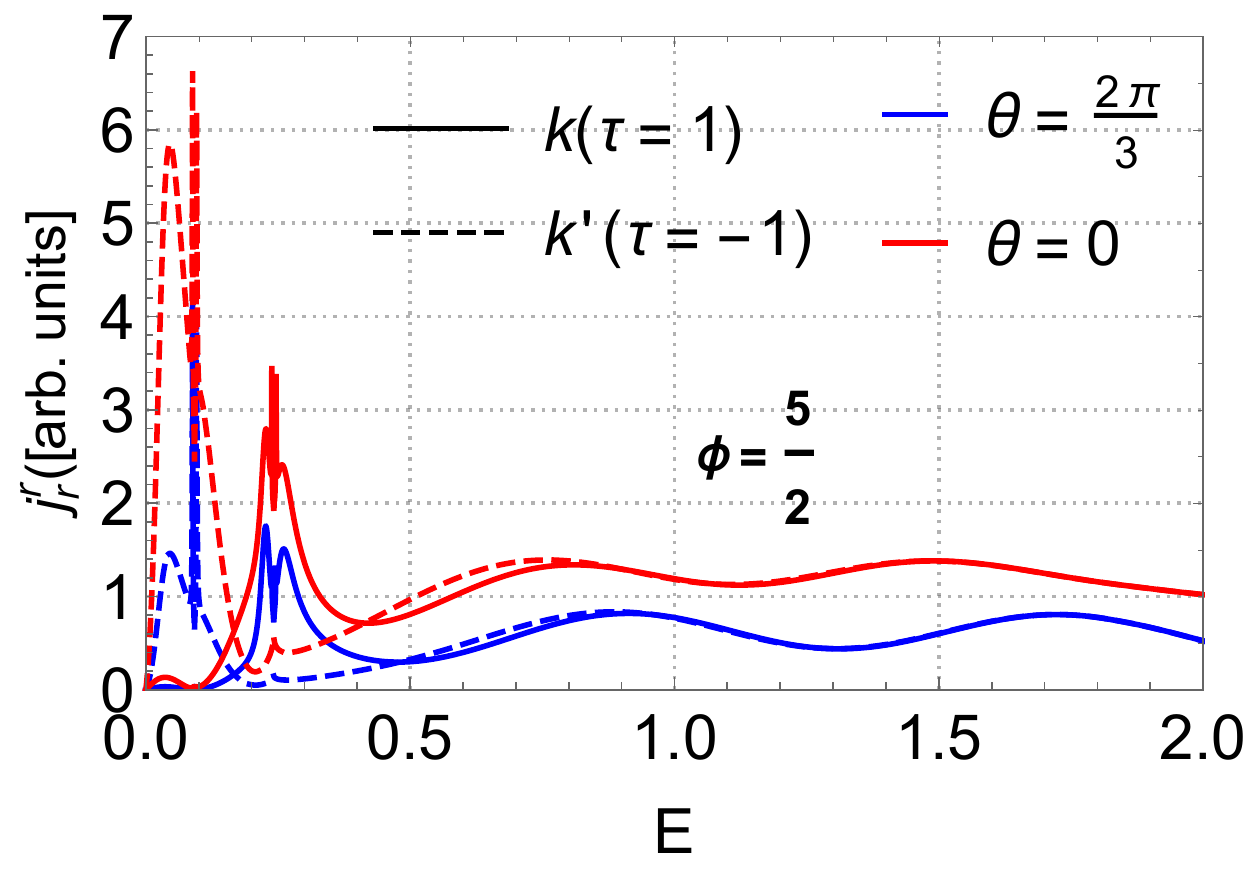}
		\label{Fig:SubFigB}
	}
	\subfloat[spin-down ($\downarrow$)]{
		\includegraphics[scale=0.62]{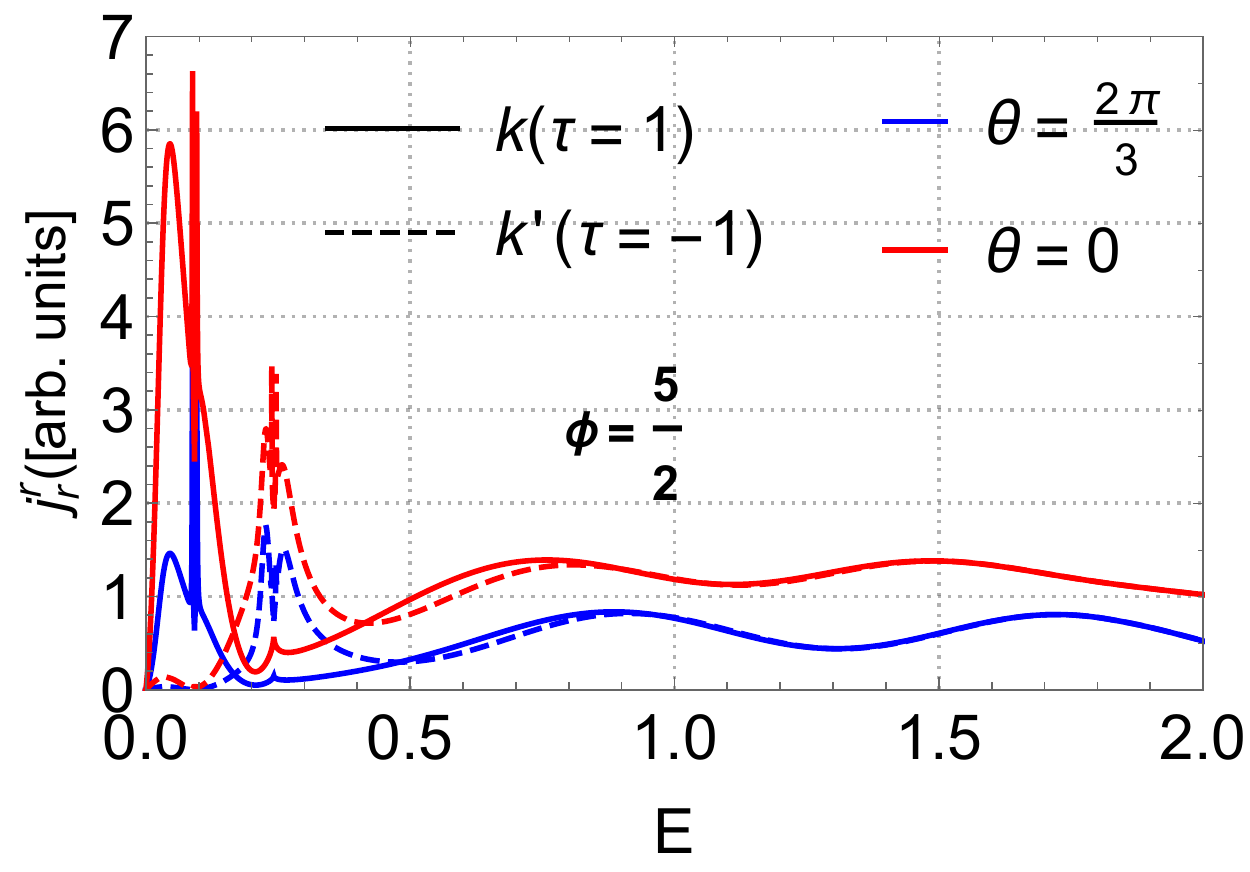}
		\label{Fig:SubFigB}
	}
	\caption{The radial component of the far-field scattered current $j_r^r$ as a function of the incident energy $E$ for the angles $\theta=2\pi/3$ (blue line) and $\theta=0$ (red line) with $V=1$ and $R=2$. (a): ($k$, $k'$, spin-up ($\uparrow$), $\phi=1/2$) and (b): ($k$, $k'$, spin-down ($\downarrow$), $\phi=1/2$) and (c): ($k$, $k'$, spin-up ($\uparrow$),$\phi=5/2$)and (d) : ($k$, $k'$, spin-down ($\downarrow$), $\phi=5/2$) states.}
	\label{fig6}
\end{figure}

Figure \ref{fig6} shows the radial component of the far-field scuttering current $j_r^{r}$ as a function of the incident energy $E$ with $\theta=0$ (red line) and $\theta=2\pi/3$ (blue line) for the states ($\uparrow$, $\pm\tau$, $\phi=3/2$) (\ref{fig6}(a) and  \ref{fig6}(b)) and  ($\uparrow$, $\pm\tau$, $\phi=5/2$) [\ref{fig6} (c) and  \ref{fig6} (d)] with $V=1$ and $R=4$.
In Figure \ref{fig6}(a) and \ref{fig6}(b) it is shown that when $E\rightarrow 0$, $j_r^{r}\rightarrow 0$, for both values of $\theta$, as $E$ increases to the value 0.1, we notice the appearance of resonance peaks with a maximum peak for the states ($\uparrow$,$-\tau$, $\theta=0$) and ($\downarrow$,$\tau$, $\theta=0$) respectively, therefore in the regime $0.1< E < 1$, $j_r^{r}$ possesses an oscillatory behavior, however, in the regime $E>1$, $j_r^{r}$ exhibits damped oscillatory behavior with the existence of the symmetry
$j_r^{r}(\uparrow, \tau, \theta)=j_r^{r}(\uparrow, -\tau, \theta)$ and $j_r^{r}(\downarrow, \tau, \theta)=j_r^{r}(\downarrow, -\tau, \theta)$.
In Figure \ref{fig6}(c) and \ref{fig6}(d) we show the variation of $j_r^{r}$ as a function of $E$ for the states ($\uparrow$, $\pm \tau$, $\phi=5/2$) and ($\downarrow$,$\pm \tau$, $\phi=5/2$), these Figures show that the behavior of $j_r^{r}$ is similar to that of Figures \ref{fig6}(a) and \ref{fig6}(b), the only differences is that the amplitude of $j_r^{r}$ for $\phi=3/2$ is larger than the amplitude of $j_r^{r}$ for $\phi=5/2$.
\begin{figure}[]\centering
	\subfloat[spin-up ($\uparrow$)]{
		\includegraphics[scale=0.62]{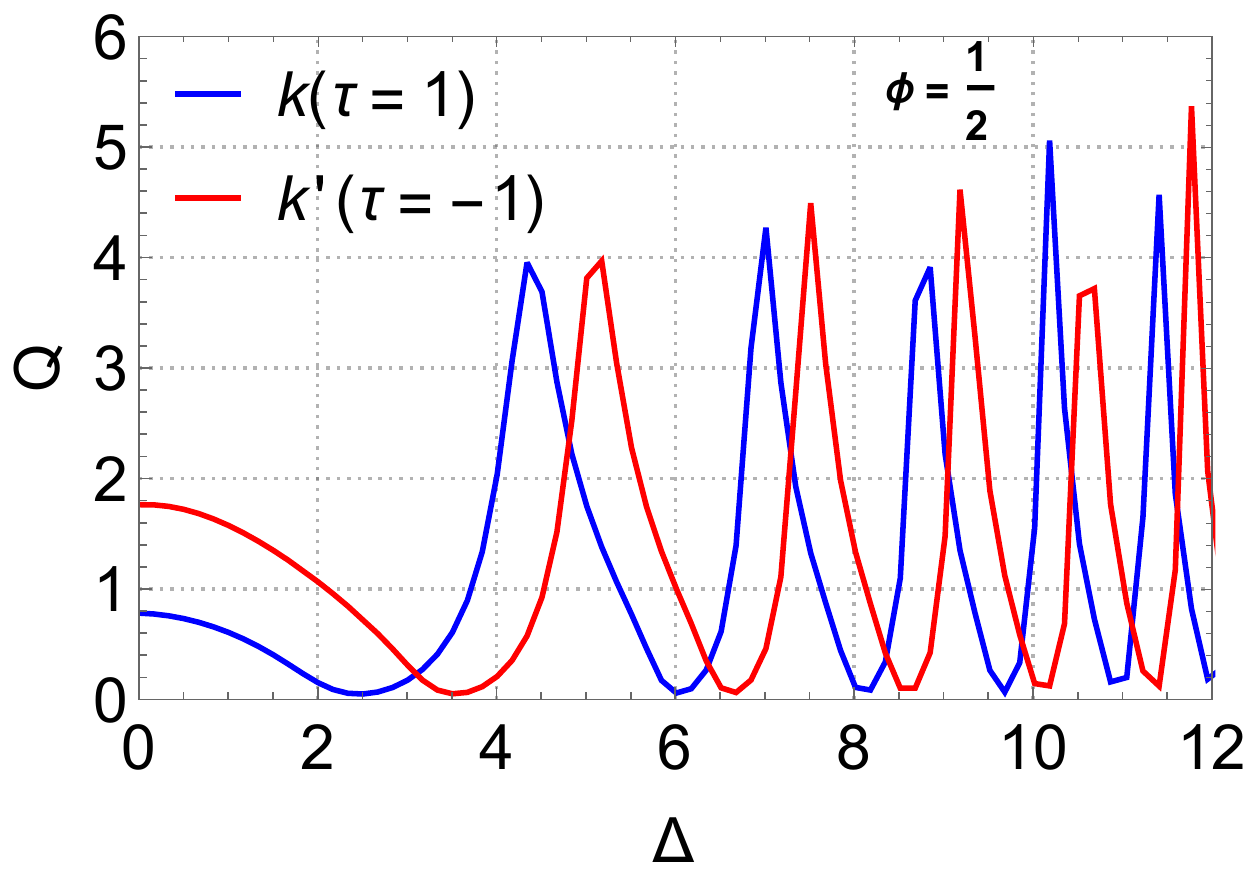}
		\label{Fig:SubFigB}
	}
	\subfloat[spin-down ($\downarrow$)]{
		\includegraphics[scale=0.62]{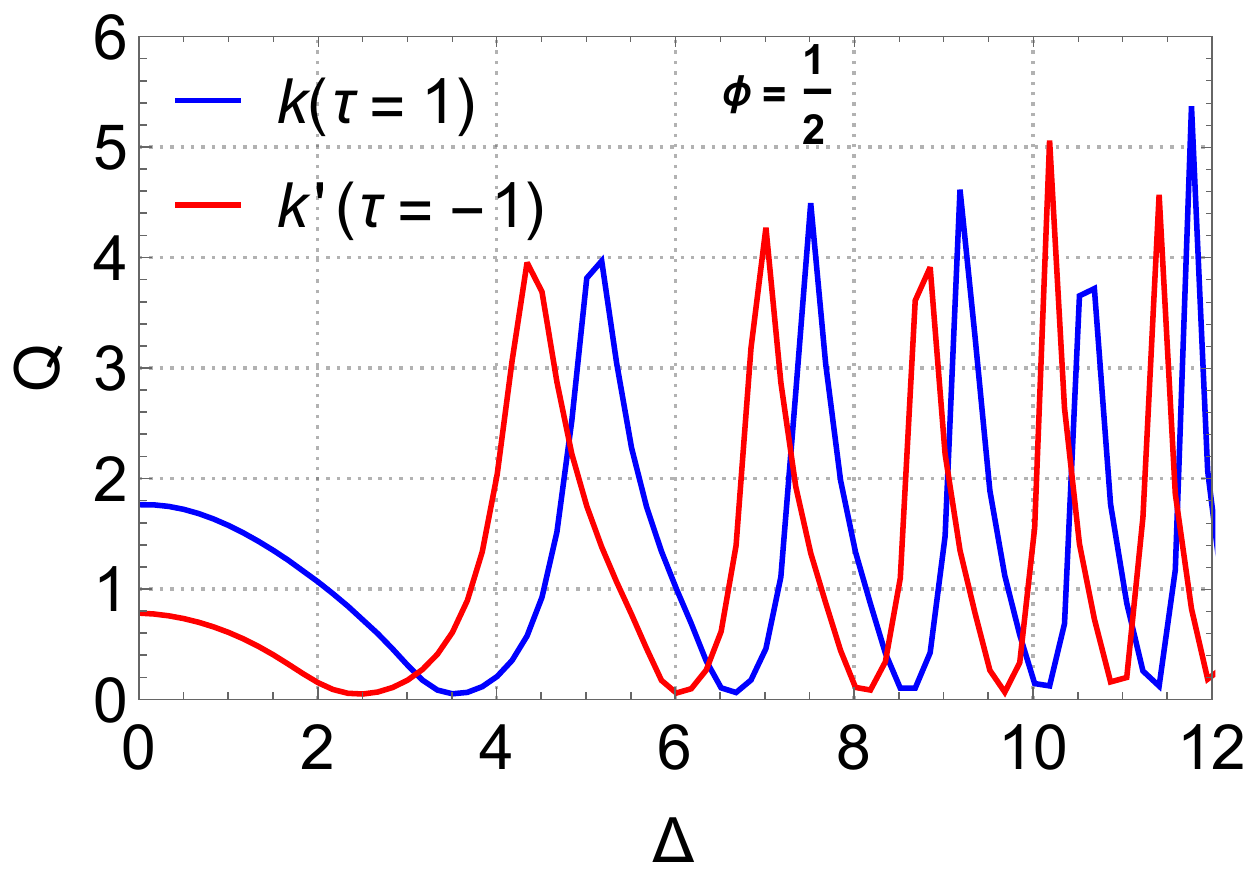}
		\label{Fig:SubFigB}
	}\newline
	\subfloat[spin-up ($\uparrow$)]{
		\includegraphics[scale=0.62]{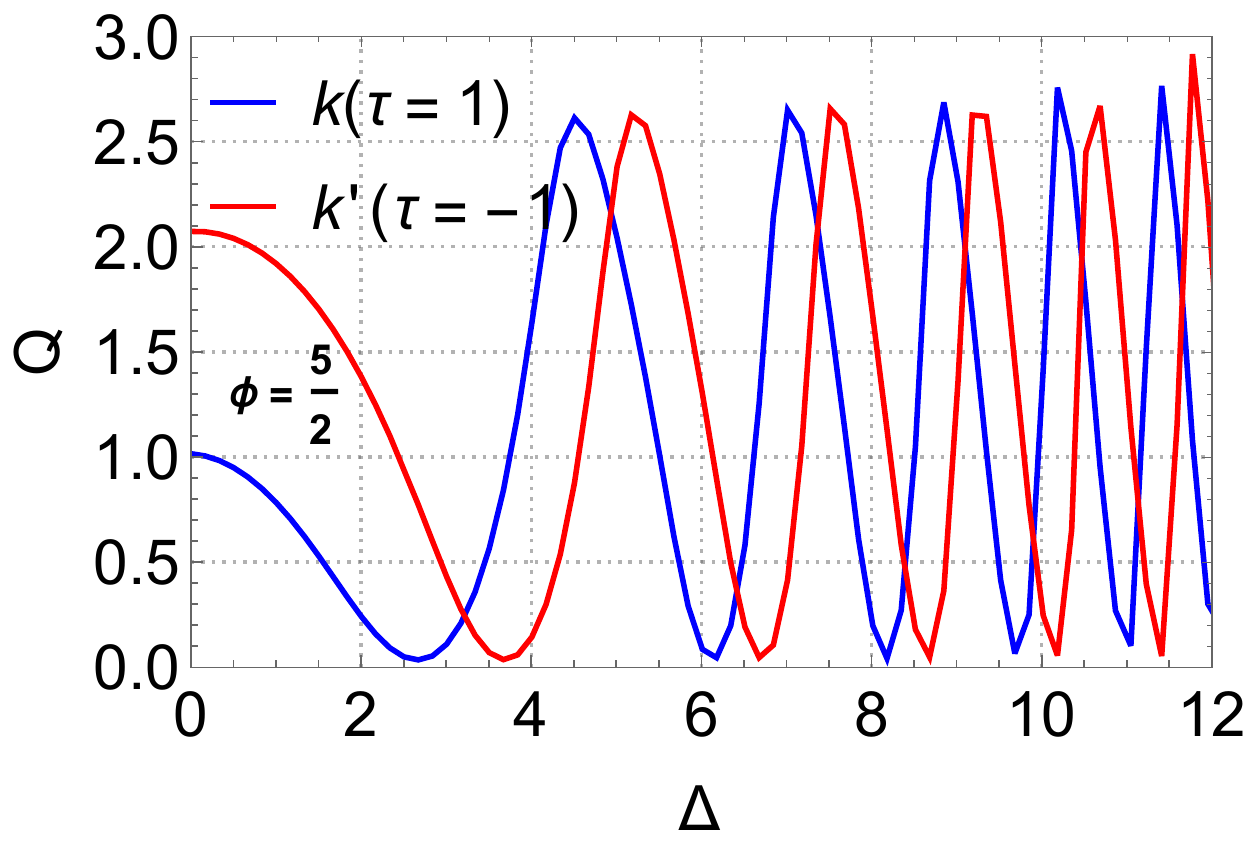}
		\label{Fig:SubFigB}
	}
	\subfloat[spin-down ($\downarrow$))]{
		\includegraphics[scale=0.62]{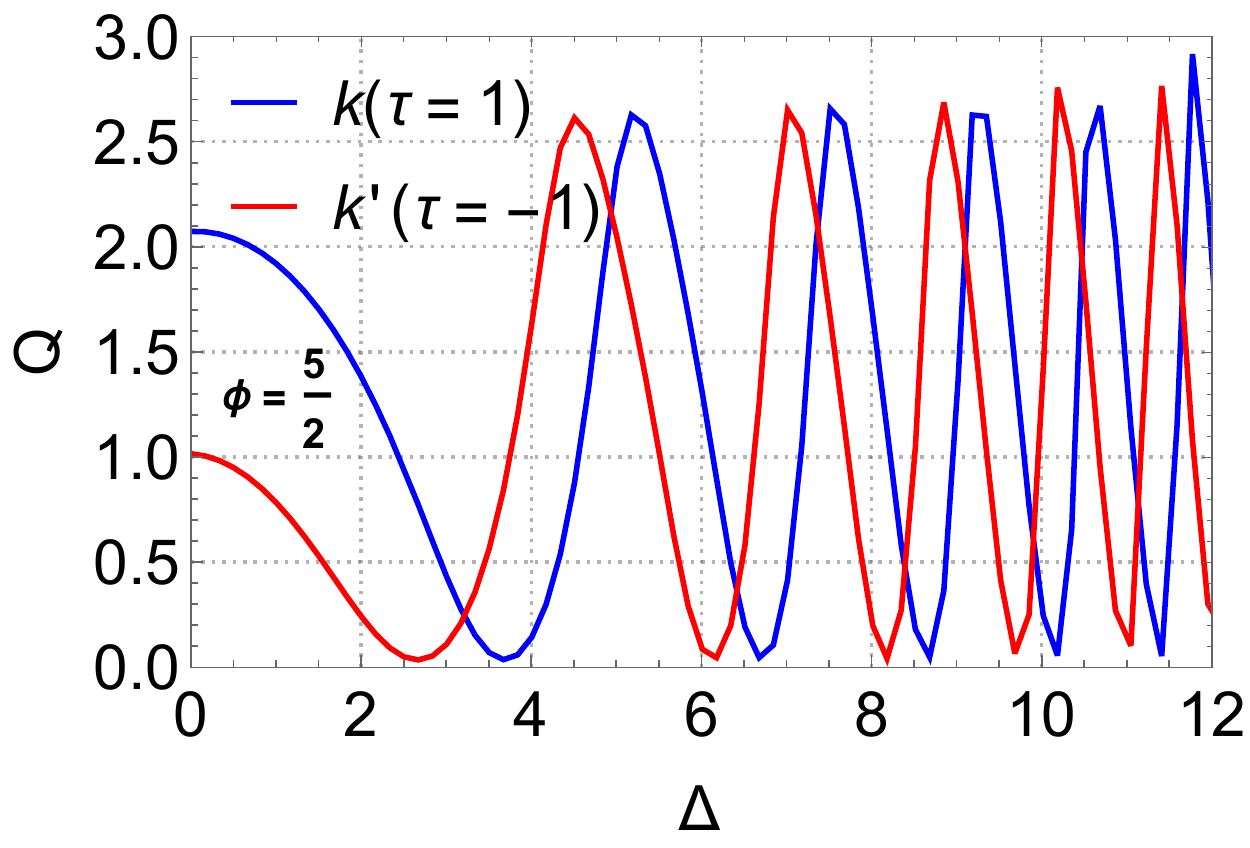}
		\label{Fig:SubFigB}
	}
	\caption{The radial component of the far-field scattered current $j_r^r$ as a function of the incident energy $E$ for the angles $\theta=2\pi/3$ (blue line) and $\theta=0$ (red line) with $V=1$ and $R=2$. (a): ($k$, $k'$, \emph{spin up}) and (b): ($k$, $k'$, \emph{spin down}) states.}
	\label{Fig61}
\end{figure}

Figure \ref{Fig61} shows the scattering efficiency $Q$ as a function of the gap $\Delta$ applied inside the quantum dot for $R=10$, $V=1$ and $E=1.2$ for the states (a): ($\uparrow$,$\pm\tau$, $\phi=1/2$)), (b): ($\downarrow$, $\pm\tau$, $\phi=1/2$), (c): ($\uparrow$, $\pm\tau$, $\phi=5/2$) and (d): ($\downarrow$,$\pm\tau$, $\phi=5/2$) with $\tau=1$(blue curve) and $\tau=-1$ (red curve).
For Figure \ref{Fig61}(a) we observe that when $\Delta\rightarrow 0$, $Q$ has fixed values with $Q(\uparrow, \tau=1, \phi=1/)=0.8$ and $Q(\uparrow, \tau=-1, \phi=1/)=1.8$, then $Q$ decreases until a minimum value depending on $\tau$ and after that it shows a continuous oscillatory behavior for both values of $\tau$, note that the amplitudes of $Q$ increase with increasing $\Delta$ respectively, on the other hand figure \ref{Fig61}(b) shows the same behavior as Figure \ref{Fig61}(a) with symmetry
$Q(\uparrow, \tau=1, \phi=1/2)=Q(\uparrow, \tau=-1, \phi=1/2)$ when the spin state $(\uparrow)\rightarrow (\downarrow)$ and $\tau \rightarrow -\tau$. In figure \ref{Fig61}(c) when $\Delta \rightarrow 0$, $Q(\tau=1)=1$ and $Q(\tau=-1)=2.2$, when $\Delta$ increases, $Q$ presents minima and maxima with an oscillatory behavior, for figure \ref{Fig61}(d) we notice that this figure has the same evolution as figure \ref{Fig61}(c) when $\tau \rightarrow -\tau$ and $(\uparrow)\rightarrow (\downarrow)$, moreover, the comparison of Figures \ref{Fig61} (a,b) for $\phi=1/2$ and \ref{Fig61} (c,d) for $\phi=5/2$ shows that when $\phi$ increases the amplitude of $Q$ decreases and vice versa is also true.
%========================================================
\section{Conclusion}\label{Conclsion}
%========================================================
We investigated the scattering of Dirac electrons in a quantum dot of $\mathrm{MoS_2}$ subjected to potential barrier V and  the magnetic flux $\phi$. The Dirac equation's proper boundary conditions have been calculated, and it has been demonstrated how the scattering coefficients $\alpha_m$ and $\beta_m$ may be used formally to describe the features of our systems. The radial component of current density, the square modulus of the scattering coefficient, and the scattering efficiency were all determined.

We investigated the Dirac electron scattering in a $\mathrm{MoS_2}$ quantum dot as a function of the quantum dot radius R for spin-up and spin-down states, we found that as $R$ approaches zero, the efficiency scattering $Q$ becomes zero, so when $R$ increases, $Q$ exhibits a damped oscillatory characteristic, with the appearance of emerging peaks depends on $\phi$ and $\tau$. Then, we established the effect of the incident energy E on the scattering phenomenon, it was shown that when E increases, Q decreases rapidly with the appearance of a single resonance peak for both spin states. Moreover, we found that Q exhibits oscillatory behavior as a function of gap energy $\Delta$ with the symmetry behavior $Q(\uparrow, \tau, \phi)=Q(\uparrow, -\tau, \phi$),
Furthermore, We have analyzed at the square modulus of the diffusion coefficients $\zeta_m$ as a function of incident energy to find the resonances, we have clearly shown that near $E=0$ all $\zeta_m$ cancel out except the lowest $\zeta_0$, as we increase $E$, the contribution of higher order scattering coefficients ($m=1, 2, 3$) begin to appear. Additionally, for larger energy, the $\zeta^2_m$ tend toward oscillatory behaviour. Regarding the angular characteristic of the reflected radial component, we have found that each mode has ($2m+1$) preferred directions of scattering observable with different amplitudes.
%=======================================================

%========================================================

\begin{thebibliography}{99}
%=======================================================
\bibitem{Novoselov04} K. S. Novoselov, A. K. Geim, S. V.  Morozov, D. Jiang, Y. Zhang, S. V.  Dubonos, I. V. Gregorieva  and A. A.  Firsov  Science 306, (2004).
%
\bibitem{Novoselov041} K. S. Novoselov , D.  Jiang , F.  Schedin, T. J.  Booth ,  V. V. Khotkivich , S. V. Morozov  and A. K.  Geim , Proc.  Natl. Acad. Sci. USA 102, 10451 (2004).
%
\bibitem{Geim13} A. K.  Geim and  I. V Grigorieva Nature 499,  419 (2013).
%
\bibitem{Wang12}  Q. H. Wang, K. K. Zadeh,  A. Kis, J. n.  Coleman  and M. S. Strano  Nature Nanotech. 7, 699 (2012).
%
\bibitem{Yazyev15} O. V. Yazyev and  A. Kis  Materials Today 18,  20 (2015).
%
\bibitem{Radisavljevic01}  B. Radisavljevic,  A. Radenovic, J.  Brivio , V.  Giacometti  and A.  Kis   Nature Nanotech. 6, 147 (2001).
%
\bibitem{Mak10} K. F. Mak ,  C. Lee C,  J. Hone ,  J. Shan  and T. F.  Heinz Phys. Rev. Lett. 105, 136805 (2010).
%
\bibitem{Nayak12} A. P. Nayak, S.  Bhattacharyya ,  J. Zhu,  J. Liu , X. Wu ,  T. Pandey , C.  Jin,  A. K. Singh , D.  Akinwande and J. F. Lin, Nature Comm. 5 3731 (2012).
%
\bibitem{Kang12}[10]  Q. Y. J. Kang,  Z. Shao, X.  Zhang,  S. Chang,  G. Wang,  S. Qin and  J. Li,  Phys. Lett. A 376, 1166 (2012).
%
\bibitem{Conley13} H. J.  Conley, B. Wang, J. I.  Ziegler,  J. R. F. Haglund Jr,  S. T. Pantelides  and  K. I. Bolotin , Nano Lett. 13, 3626 (2013).
%
\bibitem{Radisavljevic13}  B. Radisavljevic  and  A. Kis ,  Nature Mater. 12, 815 (2013).
%
\bibitem{Xiao12}  D. Xiao ,  G. B. Liu , W. Feng, X. Xu  and W. Yao, Phys. Rev. Lett. 108, 196802 (2012).
%
\bibitem{Kadantsev12} E. S. Kadantsev and  P. Hawrylak  Solid State Comm. 152, 909 (2012).
%
\bibitem{Ping17} J. Ping, Z. Fan, M. Sindoro, Y. Ying, H. Zhang , Adv. Funct. Mater. 27, 1605817 (2017).

\bibitem{Singh17} E. Singh, K.S. Kim, G.Y. Yeom, H.S. Nalwa,  RSC Adv. 7, 28234 (2017).

\bibitem{Tan15} C. Tan, H. Zhang,  Chem. Soc. Rev. 44, 2713 (2015).

\bibitem{Gong17} L. Gong, L. Yan, R. Zhou, J. Xie, W. Wu, Z. Gu,  J. Mater. Chem. B 5, 1873 (2017).
%
\bibitem{Yadav19} V. Yadav, S. Roy, P. Singh, Z. Khan, A. Jaiswal,  15, 1803706 (2019).
%
\bibitem{Arul16} N.S. Arul, V.D. Nithya,  RSC Adv. 6, 65670 (2016).
%
\bibitem{Heinl13} J. Heinl, M. Schneider and P. W. Brouwer,  Phys. Rev. B 87, 245426 (2013).
%
\bibitem{Belouad18} A. Belouad A, Y. Zahidi, A. Jellal  and H. Bahlouli, Europhys. Lett. 123, 28002 (2018).
%
\bibitem{Oliveira16} D. Oliveira, Jiyong Fu, L. Villegas-Lelovsky, A. C. Dias, and Fanyao Qu,  Phys. Rev. B 93, 205422 (2016).
%
\bibitem{Di Xiao12} D. Xiao, G. B. Liu, W. Feng, X. Xu, and W. Yao, Phys. Rev. Lett  108, 196802 (2012).
%
\bibitem{Schnez08} S. Schnez, K. Ensslin, M. Sigrist, and T. Ihn, Phys. Rev. B 78, 195427 (2008).

\bibitem{Heinisch13} R. L. Heinisch, F. X. Bronold, and H. Fehske, Phys. Rev. B 87, 155409 (2013).

\bibitem{Berry13} M. V. Berry and R. J. Mondragon, Proc. R. Soc. London A 412, 53 (1987).

\bibitem{Grujic11} M. Grujic, M. Zarenia, A. Chaves, M. Tadie, G. A. Farias, and F, M. Peeters, Phys. Rev. B 84, 205441 (2011).
%
\bibitem{Schulz15} C. Schulz, R. L. Heinisch, and H. Fehske, Phy. Rev. B 91, 045130 (2015).
%
\bibitem{Zheng19} Z. Xue Wu, C. Y. Long, J. Wu, Z. Wen Long and T. Xu, Int. J. Mod. Phys. A, 4 1950212 (2019).
%========================================================
\end{thebibliography}
\end{document}